

\input harvmac

\noblackbox
\baselineskip 20pt plus 2pt minus 2pt

\overfullrule=0pt


\def \eq#1 {\eqno {(#1)}}

\def\bs{\bigskip}

\def\hb{\hfill\break}
\def\qq{\qquad}
\def\bl{\bigl}
\def\br{\bigr}

\def\IR{\relax{\rm I\kern-.18em R}}
\def \ra {\rightarrow}

\def\np {  Nucl. Phys. }
\def \pl { Phys. Lett. }

\def \prl { Phys. Rev. Lett. }
\def \pr  { Phys. Rev. }
\def \prd  { Phys. Rev. }


\def\r{\rho}
\def\a{\alpha}
\def\A{\Alpha}
\def\b{\beta}

\def\g{\gamma}

\def\d{\delta}
\def\D{\Delta}
\def\e{\epsilon}

\def\P{\Phi}

\def\th{\theta}

\def\m{\mu}
\def\n{\nu}

\def\Om{\Omega}
\def\l{\lambda}

\def\s{\sigma}

\def\cA{{\cal A}}

\def\IR{\relax{\rm I\kern-.18em R}}

\def \bd {\bar \del}

\def \z { {\bar z} }
\def \A { {\bar A} }

\def \ha {{1\over 2}}

\def \ov {\over}
\def\ints{\int d^2 z}
\def\diag{{\rm diag}}
\def\const{{\rm const.}}

\def \cA {{\cal A}}


\lref\BSthree{I. Bars and K. Sfetsos, Mod. Phys. Lett. {\bf A7} (1992) 1091.}

\lref\BShet{I. Bars and K. Sfetsos, Phys. Lett. {\bf 277B} (1992) 269.}

\lref\BSglo{I. Bars and K. Sfetsos, Phys. Rev. {\bf D46} (1992) 4495.}

\lref\BSexa{I. Bars and K. Sfetsos, Phys. Rev. {\bf D46} (1992) 4510.}

\lref\SFET{K. Sfetsos, Nucl. Phys. {\bf B389} (1993) 424.}

\lref\BSslsu{I. Bars and K. Sfetsos, Phys. Lett. {\bf 301B} (1993) 183.}

\lref\BSeaction{I. Bars and K. Sfetsos, Phys. Rev. {\bf D48} (1993) 844.}

\lref\BN{ I. Bars and D. Nemeschansky, Nucl. Phys. {\bf B348} (1991) 89.}

\lref\WIT{E. Witten, Phys. Rev. {\bf D44} (1991) 314.}

 \lref\IBhet{ I. Bars, Nucl. Phys. {\bf B334} (1990) 125. }

 \lref\IBCS{ I. Bars, ``String Propagation on Black Holes'', USC-91-HEP-B3.\hb
{\it Curved Space-time Strings and Black Holes},
in Proc.
 {\it XX$^{th}$ Int. Conf. on Diff. Geometrical Methods in Physics}, eds. S.
 Catto and A. Rocha, Vol. 2, p. 695, (World Scientific, 1992).}

 \lref\CRE{M. Crescimanno, Mod. Phys. Lett. {\bf A7} (1992) 489.}

\lref\clapara{K. Bardakci, M. Crescimanno and E. Rabinovici,
Nucl. Phys. {\bf B344} (1990) 344.}

\lref\MSW{G. Mandal, A. Sengupta and S. Wadia,
Mod. Phys. Lett. {\bf A6} (1991) 1685.}

 \lref\HOHO{J.B. Horne and G.T. Horowitz, Nucl. Phys. {\bf B368} (1992) 444.}

 \lref\FRA{E. S. Fradkin and V. Ya. Linetsky, Phys. Lett. {\bf 277B}
          (1992) 73.}

 \lref\ISH{N. Ishibashi, M. Li  and A. R. Steif,
         Phys. Rev. Lett. {\bf 67} (1991) 3336.}

 \lref\HOR{P. Horava, Phys. Lett. {\bf 278B} (1992) 101.}

 \lref\RAI{E. Raiten, ``Perturbations of a Stringy Black Hole'',
         Fermilab-Pub 91-338-T.}

 \lref\GER{D. Gershon, Phys. Rev. {\bf D49} (1994) 999.}

\lref\GERexa{D. Gershon, ``Semiclassical vs. Exact Solutions of charged Black
Hole in four dimensions and exact $O(D,D)$ duality'', TAUP-2121-93,
hep-th/9311122.}

\lref\GERexadual{D. Gershon, ``Exact $O(D,D)$ transformations in WZW models",
TAUP-2129-93, hep-th/9312154.}

 \lref \GIN {P. Ginsparg and F. Quevedo,  Nucl. Phys. {\bf B385} (1992) 527. }

 \lref\HOHOS{ J.H. Horne, G.T. Horowitz and A. R. Steif, Phys. Rev. Lett.
 {\bf 68} (1991) 568.}

 \lref\groups{
 M. Crescimanno. Mod. Phys. Lett. {\bf A7} (1992) 489. \hb
 J. B. Horne and G.T. Horowitz, Nucl. Phys. {\bf B368} (1992) 444. \hb
 E. S. Fradkin and V. Ya. Linetsky, Phys. Lett. {\bf 277B} (1992) 73. \hb
 P. Horava, Phys. Lett. {\bf 278B} (1992) 101.\hb
 E. Raiten, ``Perturbations of a Stringy Black Hole'',
         Fermilab-Pub 91-338-T.\hb
 D. Gershon, ``Exact Solutions of Four-Dimensional Black Holes in
         String Theory'', TAUP-1937-91.}

\lref\NAWIT{C. Nappi and E. Witten, Phys. Lett. {\bf 293B} (1992) 309.}

\lref\FRATSE{E. S. Fradkin and A.A. Tsey,
Phys. Lett. {\bf 158B} (1985) 316.}

\lref\CALLAN{ C.G. Callan, D. Friedan, E.J. Martinec and M. Perry,
Nucl. Phys. {\bf B262} (1985) 593.}

\lref\DB{L. Dixon, J. Lykken and M. Peskin, Nucl. Phys.
{\bf B325} (1989) 325.}

\lref\IB{I. Bars, Nucl. Phys. {\bf B334} (1990) 125.}

\lref\BUSCHER{T. Buscher, Phys. Lett. {\bf 194B} (1087) 59;
Phys. Lett. {\bf 201B} (1988) 466.}

\lref\RV{M. Ro\v cek and E. Verlinde, Nucl. Phys. {\bf B373} (1992) 630.}
\lref\GR{A. Giveon and M. Ro\v cek, Nucl. Phys. {\bf B380} (1992) 128. }

\lref\nadual{X. C. de la Ossa and F. Quevedo,
Nucl. Phys. {\bf B403} (1993) 377.}

\lref\duearl{B.E. Fridling and A. Jevicki,
Phys. lett. {\bf B134} (1984) 70.\hb
E.S. Fradkin and A.A. Tseytlin, Ann. Phys. {\bf 162} (1985) 31.}

\lref\ABEN{I. Antoniadis, C. Bachas, J. Ellis and D.V. Nanopoulos,
Phys. Lett. {\bf B211} (1988) 393.}

\lref\Kalofour{N. Kaloper, Phys. Rev. {\bf D48} (1993) 4658.}

\lref\GPS{S.B. Giddings, J. Polchinski and A. Strominger,
Phys. Rev. {\bf D48} (1993) 5784.}

\lref\SENrev{A. Sen, ``Black Holes and Solitons in String Theory'',
TIFR-TH-92-57.}

\lref\TSEd{A.A. Tseytlin, Mod. Phys. Lett. {\bf A6} (1991) 1721.}

\lref\TSESC{A. S. Schwarz and A.A. Tseytlin, ``Dilaton shift under duality
and torsion of elliptic complex'', IMPERIAL/TP/92-93/01. }

\lref\Dualone{K. Meissner and G. Veneziano,
Phys. Lett. {\bf B267} (1991) 33;
Mod. Phys. Lett. {\bf A6} (1991) 3397. \hb
M. Gasperini and G. Veneziano, Phys. Lett. {\bf 277B} (1992) 256. \hb
M. Gasperini, J. Maharana and G. Veneziano, Phys. Lett. {\bf B296} (1992) 51.}

\lref \cosm {G. Veneziano, \pl {\bf B265} (1991)287. \hb
M. Gasperini, J. Maharana and G. Veneziano, Phys. Lett. {\bf B296} (1992) 51.}

\lref\rovr{K. Kikkawa and M. Yamasaki, Phys. Lett. {\bf B149} (1984) 357.\hb
N. Sakai and I. Senda, Prog. theor. Phys. {\bf 75} (1986) 692.}

\lref\narain{K.S. Narain, Phys. Lett. {\bf B169} (1986) 369.\hb
K.S. Narain, M.H. Sarmadi and C. Vafa, Nucl. Phys. {\bf B288} (1987) 551.}

\lref\GV{P. Ginsparg and C. Vafa, Nucl. Phys. {\bf B289} (1987) 414.}
\lref\nssw{V. Nair, A. Shapere, A. Strominger and F. Wilczek,
Nucl. Phys. {\bf B287} (1987) 402.}

\lref\vafa{C. Vafa, ``Strings and Singularities'', HUTP-93/A028.}

\lref\Dualtwo{A. Sen,
Phys. Lett. {\bf B271} (1991) 295;\ ibid. {\bf B274} (1992) 34;
Phys. Rev. Lett. {\bf 69} (1992) 1006. \hb
S. Hassan and A. Sen, Nucl. Phys. {\bf B375} (1992) 103. \hb
J. Maharana and J. H. Schwarz, Nucl. Phys. {\bf B390} (1993) 3.\hb
A. Kumar, Phys. Lett. {\bf B293} (1992) 49.}

\lref\dualmargi{S.F. Hassan and A. Sen, Nucl. Phys. {\bf B405} (1993) 143.\hb
M. Henningson and C. Nappi, Phys. Rev. {\bf D48} (1993) 861.}

\lref\bv{R. Brandenberger and C. Vafa, Nucl. Phys. {\bf B316} (1989) 301.}
\lref\gsvy{B.R. Greene, A. Shapere, C. Vafa and S.T. Yau,
Nucl. Phys. {\bf B337} (1990) 1.}
\lref\tv{A.A. Tseytlin and C. Vafa,  Nucl. Phys. {\bf B372} (1992) 443.}

\lref\grvm{A. Shapere and F. Wilczek, Nucl. Phys. {\bf B320} (1989) 609.\hb
A. Giveon, E. Rabinovici and G. Veneziano,
Nucl. Phys. {\bf B322} (1989) 167.\hb
A. Giveon, N. Malkin and E. Rabinovici, Phys. Lett. {\bf B238} (1990) 57.}

\lref\GRV{M. Gasperini, R. Ricci and G. Veneziano,
Phys. Lett. {\bf B319} (1993) 438.}

\lref\KIRd{E. Kiritsis, Nucl. Phys. {\bf B405} (1993) 109.}

\lref\gpr{A. Giveon, M. Porrati and E. Rabinovici, ``Target Space Duality
in String Theory'', RI-1-94, hep-th/9401139.}

\lref\slt{A. Giveon, Mod. Phys. Lett. {\bf A6} (1991) 2843.}

\lref\GIPA{A. Giveon and A. Pasquinucci, ``On cosmological string backgrounds
with toroidal isometries'', IASSNS-HEP-92/55, August 1992.}

\lref\KASU{Y. Kazama and H. Suzuki, Nucl. Phys. {\bf B234} (1989) 232. \hb
Y. Kazama and H. Suzuki Phys. Lett. {\bf 216B} (1989) 112.}

\lref\WITanom{E. Witten, Comm. Math. Phys. {\bf 144} (1992) 189.}

\lref\WITnm{E. Witten, Nucl. Phys. {\bf B371} (1992) 191.}

\lref\IBhetero{I. Bars, Phys. Lett. {\bf 293B} (1992) 315.}

\lref\IBerice{I. Bars, {\it Superstrings on Curved Space-times}, Lecture
delivered at the Int. workshop on {\it String Quantum Gravity and Physics
at the Planck Scale}, Erice, Italy, June 1992.}

\lref\DVV{R. Dijkgraaf, E. Verlinde and H. Verlinde, Nucl. Phys. {\bf B371}
(1992) 269.}

\lref\TSEY{A.A. Tseytlin, Phys. Lett. {\bf 268B} (1991) 175.}

\lref\JJP{I. Jack, D. R. T. Jones and J. Panvel,
          Nucl. Phys. {\bf B393} (1993) 95.}

\lref\BST { I. Bars, K. Sfetsos and A.A. Tseytlin, unpublished. }

\lref\TSEYT{ A.A. Tseytlin, Nucl. Phys. {\bf B399} (1993) 601.}

\lref\TSEYTt{A.A. Tseytlin, Nucl. Phys. {\bf B411} (1993) 509.}

 \lref\SHIF { M. A. Shifman, Nucl. Phys. {\bf B352} (1991) 87.}
\lref\SHIFM { H. Leutwyler and M. A. Shifman, Int. J. Mod. Phys. {\bf
A7} (1992) 795. }

\lref\POLWIG { A. M. Polyakov and P. B. Wiegman, Phys.
Lett. {\bf 141B} (1984) 223.  }

\lref\BCR{K. Bardakci, M. Crescimanno
and E. Rabinovici, Nucl. Phys. {\bf B344} (1990) 344. }

\lref\Wwzw{E. Witten, Commun. Math. Phys. {\bf 92} (1984) 455.}

\lref\GKO{P. Goddard, A. Kent and D. Olive, Phys. Lett. {\bf B152} (1985) 88.}

\lref\Toda{A. N. Leznov and M. V. Saveliev, Lett. Math. Phys. {\bf 3} (1979)
489. \hb A. N. Leznov and M. V. Saveliev, Comm. Math. Phys. {\bf 74}
(1980) 111.}

\lref\GToda{J. Balog, L. Feh\'er, L. O'Raifeartaigh, P. Forg\'acs and A. Wipf,
Ann. Phys. (New York) {\bf 203} (1990) 76; Phys. Lett. {\bf 244B}
(1990) 435.}

\lref\GWZW{ E. Witten, \np {\bf B223} (1983) 422. \hb
K. Bardakci, E. Rabinovici and B. S\"aring, Nucl. Phys. {\bf B299}
(1988) 157. \hb K. Gawedzki and A. Kupiainen, Phys. Lett. {\bf 215B}
(1988) 119.; Nucl. Phys. {\bf B320} (1989) 625. }

\lref\SCH{ D. Karabali, Q-Han Park, H. J. Schnitzer and Z. Yang,
                   Phys. Lett. {\bf B216} (1989) 307. \hb D. Karabali
and H. J. Schnitzer, Nucl. Phys. {\bf B329} (1990) 649. }

 \lref\KIR{E. Kiritsis, Mod. Phys. Lett. {\bf A6} (1991) 2871. }

\lref\BIR{N. D. Birrell and P. C. W. Davies,
{\it Quantum Fields in Curved Space}, Cambridge University Press.}

\lref\WYB{B. G. Wybourn, {\it Classical Groups for Physicists }
(John Wiley \& sons, 1974).}

\lref\Brinkman{H.W. Brinkmann, Math. Ann. {\bf 94} (1925) 119.}

\lref\SANTAa{R. G\"uven, Phys. Lett. {\bf B191} (1987) 275.}
\lref\SANTAb{D. Amati and C. Klim\v c\'\i k, Phys. Lett. {\bf B219} (1989)
443.\hb
G. Horowitz and A.  Steif, Phys. Rev. Lett. {\bf 64} (1990) 260;
Phys. Rev. {\bf D42} (1990)1950.}
\lref\SANTAc{R.E. Rudd, Nucl. Phys. {\bf B352} (1991) 489.\hb
C. Duval, G.W. Gibbons and P.A. Horvathy, Phys. Rev. {\bf D43} (1991) 3907.}
\lref\Duval{C. Duval, Z. Horvath and P.A. Horvathy, Phys. Lett. {\bf B313}
(1993) 10.}
\lref\ber{E.A. Bergshoeff, R. Kallosh and T. Ortin, Phys. Rev. {\bf D47} (1993)
5444.}
\lref\SANT{J. H. Horne, G.T. Horowitz and A. R. Steif,
Phys. Rev. Lett. {\bf 68} (1991) 568.}

\lref\tsecov{A.A. Tseytlin, Nucl. Phys. {\bf B390} (1993) 153;
Phys. Rev. {\bf D47} (1993) 3421.}

\lref\garriga{J. Garriga and E. Verdaguer, Phys. Rev. {\bf D43} (1991) 391.}
\lref\jonu{O. Jofre and C. Nunez, ``Strings in plane wave backgrounds
revisited'', hepth/9311187.}

\lref\PRE{J. Prescill, P. Schwarz, A. Shapere, S. Trivedi and F. Wilczek,
Mod. Phys Lett. {\bf A6} (1991) 2353.\hb
C. Holzhey and F. Wilczek, Nucl. Phys. {\bf B380} (1992) 447.}

\lref\HAWK{J. B. Hartle and S. W. Hawking Phys. Rev. {\bf D13} (1976) 2188.\hb
S. W. Hawking, Phys. Rev. {\bf D18} (1978) 1747.}

\lref\HAWKI{S. W. Hawking, Comm. Math. Phys. {\bf 43} (1975) 199.}

\lref\HAWKII{S. W. Hawking, Phys. Rev. {\bf D14} (1976) 2460.}

\lref\euclidean{S. Elitzur, A. Forge and E. Rabinovici,
Nucl. Phys. {\bf B359} (1991) 581. }

\lref\ITZ{C. Itzykson and J. Zuber, {\it Quantum Field Theory},
McGraw Hill (1980). }

\lref\kacrev{P. Goddard and D. Olive, Journal of Mod. Phys. {\bf A} Vol. 1,
No. 2 (1986) 303.}

\lref\BBS{F.A. Bais, P. Bouwknegt, K.S. Schoutens and M. Surridge,
Nucl. Phys. {\bf B304} (1988) 348.}

\lref\nonl{A. Polyakov, {\it Fields, Strings and Critical Phenomena}, Proc. of
Les Houses 1988, eds. E. Brezin and J. Zinn-Justin North-Holland, 1990.\hb
Al. B. Zamolodchikov, preprint ITEP 87-89. \hb
K. Schoutens, A. Sevrin and P. van Nieuwenhuizen, Proc. of the Stony Brook
Conference {\it Strings and Symmetries 1991}, World Scientific,
Singapore, 1992. \hb
J. de Boer and J. Goeree, ``The Effective Action of $W_3$ Gravity to all
\hb orders'', THU-92/33.}

\lref\HOrev{G.T. Horowitz, {\it The Dark Side of String Theory:
Black Holes and Black Strings}, Proc. of the 1992 Trieste Spring School on
String Theory and Quantum Gravity.}

\lref\HSrev{J. Harvey and A. Strominger, {\it Quantum Aspects of Black
Holes}, Proc. of the 1992 Trieste Spring School on
String Theory and Quantum Gravity.}

\lref\GM{G. Gibbons, Nucl. Phys. {\bf B207} (1982) 337.\hb
G. Gibbons and K. Maeda, Nucl. Phys. {\bf B298} (1988) 741.\hb
D. Garfinkle, G. Horowitz and A. Strominger, \pr {\bf D43} (1991) 3140. }

\lref\GID{S. B. Giddings, Phys. Rev. {\bf D46} (1992) 1347.}

\lref\PRErev{J. Preskill, {\it Do Black Holes Destroy Information?},
Proc. of the International Symposium on Black Holes, Membranes, Wormholes,
and Superstrings, The Woodlands, Texas, 16-18 January, 1992.}

\lref\tye{S-W. Chung and S. H. H. Tye, Phys. Rev. {\bf D47} (1993) 4546.}

\lref\eguchi{T. Eguchi, Mod. Phys. Lett. {\bf A7} (1992) 85.}

\lref\blau{M. Blau and G. Thompson, Nucl. Phys. {\bf B408} (1993) 345.}

\lref\HSBW{P. S. Howe and G. Sierra, Phys. Lett. {\bf 144B} (1984) 451.\hb
J. Bagger and E. Witten, Nucl. Phys. {\bf B222} (1983) 1.}

\lref\GSW{M. B. Green, J. H. Schwarz and E. Witten, {\it Superstring Theory},
Cambridge Univ. Press, Vols. 1 and 2, London and New York (1987).}

\lref\KAKU{M. Kaku, {\it Introduction to Superstrings}, Springer-Verlag, Berlin
and New York (1991).}

\lref\LSW{W. Lerche, A. N. Schellekens and N. P. Warner, {\it Lattices and
Strings }, Physics Reports {\bf 177}, Nos. 1 \& 2 (1989) 1, North-Holland,
Amsterdam.}

\lref\confrev{P. Ginsparg and J. L. Cardy in {\it Fields, Strings, and
Critical Phenomena}, 1988 Les Houches School, E. Brezin and J. Zinn-Justin,
eds, Elsevier Science Publ., Amsterdam (1989). \hb
J. Bagger, {\it Basic Conformal Field Theory},
Lectures given at 1988 Banff Summer Inst. on Particle and Fields,
Banff, Canada, Aug. 14-27, 1988, HUTP-89/A006, January 1989. }

\lref\CHAN{S. Chandrasekhar, {\it The Mathematical Theory of Black Holes},
Oxford University Press, 1983.}

\lref\KOULU{C. Kounnas and D. L\"ust, Phys. Lett. {\bf 289B} (1992) 56.}

\lref\PERRY{M. J. Perry and E. Teo, Phys. Rev. Lett. {\bf 70} (1993) 2669.\hb
P. Yi, Phys. Rev. {\bf D48} (1993) 2777.}

\lref\GiKi{A. Giveon and E. Kiritsis, Nucl. Phys. {\bf B411} (1994) 487.}

\lref\kar{S.K. Kar and A. Kumar, Phys. Lett. {\bf 291B} (1992) 246.}

\lref\NW{C. Nappi and E. Witten, Phys. Rev. Lett. {\bf 71} (1993) 3751.}

\lref\HK{M. B. Halpern and E. Kiritsis,
Mod. Phys. Lett. {\bf A4} (1989) 1373.}

\lref\MOR{A.Yu. Morozov, A.M. Perelomov, A.A. Rosly, M.A. Shifman and
A.V. Turbiner, Int. J. Mod. Phys. {\bf A5} (1990) 803.}

\lref\KK{E. Kiritsis and C. Kounnas, Phys. Lett. {\bf B320} (1994) 264.}

\lref\KST{K. Sfetsos and A.A. Tseytlin,
``Antisymmetric tensor coupling and conformal
invariance in sigma models corresponding to gauged WZNW theories'',
THU-93/25, CERN-TH.6969/93, hep-th/9310159,
to appear in Phys. Rev. {\bf D} (1994).}

\lref\KSTh{K. Sfetsos and A.A. Tseytlin, Nucl. Phys. {\bf B415} (1994) 116.}

\lref\KP{S.P. Khastgir and A. Kumar, ``Singular limits and string solutions'',
IP/BBSR/93-72, hep-th/9311048.}

\lref\etc{K. Sfetsos, Phys. Lett. {\bf B324} (1994) 335.}

\lref\KTone{C. Klim\v c\'\i k and A.A. Tseytlin, Phys. Lett. {\bf B323} (1994)
305.}

\lref\KTtwo{C. Klim\v c\'\i k and A.A. Tseytlin, ``Exact four dimensional
string
solutions and Toda-like sigma models from `null-gauged' WZNW theories",
Imperial/TP/93-94/17, hep-th/9402120.}

\lref\saletan{E.J. Saletan, J. Math. Phys. {\bf 2} (1961) 1.}
\lref\jao{D. Cangemi and R. Jackiw, Phys. Rev. Lett. {\bf 69} (1992) 233.}
\lref\jat{D. Cangemi and R. Jackiw, Ann. Phys. (NY) {\bf 225} (1993) 229.}

\lref\ORS{ D. I. Olive, E. Rabinovici and A. Schwimmer, Phys. Lett. {\bf B321}
(1994) 361.}

\lref\edc{K. Sfetsos, ``Exact String Backgrounds from WZW models based on
Non-semi-simple groups'', THU-93/31, hep-th/9311093, to appear in
Int. J. Mod. Phys. {\bf A} (1994).}

\lref\sfedual{K. Sfetsos, ``Gauged WZW models and Non-abelian Duality'',\hb
THU-94/01, hep-th/9402031.}

\lref\grnonab{A. Giveon and M. Ro\v cek, ``On non-abelian duality'',
ITP-SB-93-44, RI-152-93, hep-th/9308154.}

\lref\wigner{E. ${\rm In\ddot on\ddot u}$ and E.P. Wigner,
Proc. Natl. Acad. Sci. U. S. {\bf 39} (1953) 510.}

\lref\HY{ J. Yamron and M.B. Halpern, Nucl. Phys. {\bf B351} (1991) 333.}
\lref\HB{K. Bardakci and M.B. Halpern, Phys. Rev. {\bf D3} (1971) 2493.}
\lref\MBH{M.B. Halpern, Phys. Rev. {\bf D4} (1971) 2398.}
\lref\balog{J. Balog, L. O'Raifeartaigh, P. Forgacs and A. Wipf,
Nucl. Phys. {\bf B325} (1989) 225.}
\lref\kacm{V. G. Kac, Funct. Appl. {\bf 1} (1967) 328.\hb
R.V. Moody, Bull. Am. Math. Soc. {\bf 73} (1967) 217.}
\lref \bha { K. Bardakci and M.B. Halpern, Phys. Rev. {\bf D3} (1971) 2493.\hb
M.B. Halpern, Phys. Rev. {\bf D4} (1971) 2398.}
\lref\balog{J. Balog, L. O'Raifeartaigh, P. Forgacs and A.}

\lref\TSEma{A.A. Tseytlin, ``On a Universal class of WZW-type
conformal models'', CERN-TH.7068/93, hep-th/9311062.}

\lref\BCHma{J. de Boer, K. Clubock and M.B. Halpern, ``Linearized form
of the Generic Affine-Virasoro Action'', UCB-PTH-93/34, hep-th/9312094.}

\lref\AABL{E. \' Alvarez, L. \' Alvarez-Gaum\' e,  J.L.F. Barbon and Y. Lozano,
Nucl. Phys. {\bf B415} (1994) 71.}


\lref\noumo{N. Mohammedi, ``On Bosonic and Superconformal Current Algebra
for Non-semi-simple Groups'', BONN-HE-93-51, hep-th/9312182.}

\lref\sfetse{K. Sfetsos and A.A. Tseytlin, unpublished (February 1994).}
\lref\figsonia{J.M. Figueroa-O'Farrill and S. Stanciu,
``Non-semi-simple Sugawara constructions'', QMW-PH-94-2, hep-th/9402035.}

\lref\ALLleadord{E. Witten, Phys. Rev. {\bf D44} (1991) 314.\hb
J.B. Horne and G.T. Horowitz, Nucl. Phys. {\bf B368} (1992) 444.\hb
M. Crescimanno, Mod. Phys. Lett. {\bf A7} (1992) 489.\hb
I. Bars and K. Sfetsos, Mod. Phys. Lett. {\bf A7} (1992) 1091.\hb
E.S. Fradkin and V.Ya. Linetsky, Phys. Lett. {\bf 277B} (1992) 73. \hb
I. Bars and K. Sfetsos, Phys. Lett. {\bf 277B} (1992) 269.\hb
P. Horava, Phys. Lett. {\bf 278B} (1992) 101.\hb
E. Raiten, ``Perturbations of a Stringy Black Hole'', Fermilab-Pub 91-338-T.\hb
P. Ginsparg and F. Quevedo,  Nucl. Phys. {\bf B385} (1992) 527.\hb
more}

\lref\ALLexact{
R. Dijkgraaf, E. Verlinde and H. Verlinde, Nucl. Phys. {\bf B371} (1992) 269.
\hb
I. Bars and K. Sfetsos, Phys. Rev. {\bf D46} (1992) 4510.;
Phys. Lett. {\bf 301B} (1993) 183.\hb
K. Sfetsos, Nucl. Phys. {\bf B389} (1993) 424.\hb
Gerhon}

\lref \anton { I. Antoniadis, C. Bachas, J. Ellis and D.V. Nanopoulos,
\pl {\bf B211}(1988)393. }
\lref \horav{P. Ho\v rava, \pl {\bf B278} (1992) 101.   }
\lref\gersh {  D. Gershon, preprints TAUP-1937-91, 2033-92,
 2121-93, hep-th/9210160,
9311122. }

\lref \givpassq { A. Giveon and A. Pasquinucci, \pl {\bf B294 }(1992) 162. }
\lref \napwit { C. Nappi and E. Witten, \prl {\bf 71} (1993) 3751.}

\lref \napwi { C. Nappi and E. Witten, \pl {\bf B293 }(1992) 309.}

\lref\ginspqu   { P. Ginsparg and F. Quevedo, \np {\bf B385 }(1992) 527.}
\lref\kounn { C. Kounnas and D. L\" ust, \pl {\bf B289} (1992) 56.}
\lref \SFTS { K. Sfetsos and A.A. Tseytlin, \prd {\bf D49} (1994) 2933.}
\lref \HORS { G. Horowitz and A. Steif, \pl {\bf B258}  (1991) 91.   }
\lref \klits { C. Klim\v c\'\i k and A.A. Tseytlin, to be published. }

\lref \duff{M. Duff, B. Nilsson and C. Pope, \pl {\bf B163} (1985) 343.\hb
R. Nepomechie, \pr {\bf D33} (1986) 3670. \hb
M. Duff, B. Nilsson, N. Warner and C. Pope, \pl {\bf B171} (1986) 326.}
\lref \GRT {A. Giveon, E. Rabinovici and A.A. Tseytlin, \np {\bf B409} (1993)
339. }
\lref\AAL{E. \' Alvarez, L. \' Alvarez-Gaum\' e and Y. Lozano, ``On non-abelian
duality'', CERN-TH-7204/94, hep-th/9403155.}
\lref \tpl {A.A. Tseytlin, \pl {\bf B317} (1993) 559. }

\lref \klim { C. Klim\v c\'\i k, \pl {\bf B208} (1988) 373; Lett. Math. Phys.
{21 }(1991) 23. }
\lref \mull {M. Mueller, \np {\bf B337} (1990) 37. }
\lref \napp { M.D. McGuigan, C.R. Nappi and S.A. Yost, \np {\bf B375} (1992)
421. }
\lref \absa {I. Antoniadis, C. Bachas and A. Sagnotti, \pl {\bf B235} (1990)
255. }
\lref \john {C.V. Johnson,  ``Exact models of extremal dyonic 4D black hole
solutions of heterotic string theory", IASSNS-HEP-94/20,
hep-th/9403192. }
\lref \nels {W. Nelson, ``Kaluza-Klein black holes in string theory'',
 UCSBTH-93-10, hep-th/9312058. \hb
D.A.  Lowe and A. Strominger, ``Exact Four-Dimensional Dyonic Black Holes and
Bertotti-Robinson Spacetimes in String Theory'', UCSBTH-94-14, hep-th/9403186.}

\lref \hulw{C.M. Hull and E. Witten, \pl {\bf B160} (1985) 398. }

\lref\antob{I. Antoniadis and N.A. Obers, ``Plane Gravitational Waves in String
Theory'', CPTH-A299.0494, hepth/9403191.}

\lref \nels {W. Nelson, ``Kaluza-Klein Black Holes in String theory",
UCSBTH-93-10, hep-th/9312058. }

\lref\lost {D.A.  Lowe and A. Strominger, ``Exact four-dimensional dyonic black
holes and Bertotti-Robinson spacetimes in string theory",  UCSBTH-94-14,
   hep-th/9403186.}
\lref \hulw{C.M. Hull and E. Witten, \pl {\bf B160} (1985) 398. }

\def \fourth {{1 \over 4 }}
\def \vp {\varphi}
\def \bA {{\bar {\cA}}}
\def \sll {SL(2, \IR)}
\def \su {SU(2)}
\def \uo {U(1)}

\def \sm {$\s$-model\ }
\def \sms {$\s$-models\ }
\lref \khou {R. Khouri, \pl {\bf B259} (1991) 261; \np {\bf B387} (1992)
315.\hb
C. Bachas and E. Kiritsis,  ``Exact string theory instantons by dimensional
reduction", CERN-TH.7100/93, hep-th/9311185.}
\def \ov {\over}
\def \ger {\GERexa\ }
\lref \tset { A.A. Tseytlin,
Int. J. Mod. Phys. {\bf D1 } (1992) 223, hep-th/9203033. }
\lref \barss {I. Bars, ``Curved spacetime geometry for strings and affine
non-compact algebras", USC-93/HEP-B3, hep-th/9309042.}


\baselineskip8pt
\Title{\vbox
{\baselineskip 6pt{\hbox{THU-94/08}}{\hbox
{Imperial/TP/93-94/28}}{\hbox{hep-th/9404063}} } }
{\vbox{\centerline {  Four Dimensional Plane Wave String Solutions }
\vskip4pt
 \centerline {with Coset CFT Description}\vskip2pt
 }}
\vskip -37 true pt
\centerline  { {  Konstadinos Sfetsos\footnote {$^\dagger$} {e-mail address:
sfetsos@fys.ruu.nl} }}
 \smallskip \smallskip
\centerline {\it  Institute for Theoretical Physics, Utrecht University }
\centerline {\it Princetonplein 5, TA 3508, The Netherlands}
\medskip
\centerline {and}
\medskip
\centerline{   A.A. Tseytlin\footnote{$^{*}$}{\baselineskip8pt
e-mail address: tseytlin@ic.ac.uk}\footnote{$^{\star}$}{\baselineskip5pt
On leave  from Lebedev  Physics
Institute, Moscow, Russia.} }

\smallskip\smallskip
\centerline {\it  Theoretical Physics Group, Blackett Laboratory}
\centerline {\it  Imperial College,  London SW7 2BZ, U.K. }
\medskip
\centerline {\bf Abstract}
\smallskip
\baselineskip6pt
\noindent
We present a number of  $D=4$ bosonic and heterotic string solutions with a
covariantly constant null Killing vector which, like the solution of Nappi and
Witten (NW),  correspond to  (gauged) WZW models and thus  have a direct
conformal field theory interpretation.  A class of exact plane wave solutions
(which includes the  NW solution) is constructed  by   `boosting'  the
twisted products of two $D=2$ `cosmological' or `black-hole' cosets  related
to  $(G\otimes G')/(H\otimes H')$ ($G,G'= SL(2,\IR)$ or $SU(2)$; $\  H,H'=
SO(1,1) $
or $SO(2)$)   gauged WZW models.
We describe a general limiting procedure by which one can  construct new
solutions with a covariantly constant null Killing vector  starting with known
string  backgrounds.  By applying  a  non-abelian duality transformation to the
NW model
we  find  a  $D=4$ solution which  admits
a covariantly constant null Killing vector
 but is not a plane wave. Higher dimensional bosonic  backgrounds  with
isometries can be interpreted as lower dimensional backgrounds with extra gauge
fields. Some of them are  at the same time solutions of the heterotic string
theory.
In particular, the NW  model represents also a $D=3$  gravi-electromagnetic
heterotic string plane wave.
In addition to the (1,1) supersymmetric embeddings of bosonic
solutions we construct a number of non-trivial (1,0) supersymmetric
exact $D=4$ heterotic string plane wave solutions some of which
are related (by a boost and analytic continuation) to limiting
cases of $D=4$ heterotic black hole solutions.

\Date {April 1994}

\noblackbox
\baselineskip 20pt plus 2pt minus 2pt


\vfill\eject


\def\u{{u\ov 2}}


\newsec{ Introduction }

 In order to be able to study various issues (e.g. singularities) of
gravitational physics in string theory at least to first order in the string
coupling expansion
one needs  exact  (in $\a'$) classical solutions of string equations
which admit  an explicit conformal field theory (CFT) interpretation.
 Knowing  the corresponding    CFT not only makes  possible in principle to
determine the operator content and  scattering amplitudes  but is also  crucial
in order  to give an adequate `stringy'  space-time interpretation
to  a   string solution
which  should go  beyond the naive picture of a background parametrised just by
a  set of `massless' fields ($G_{\m\n}, B_{\m\n}, \P$).

The only two classes of $exact$  classical solutions which are known at present
in bosonic string theory are the plane-wave type backgrounds
(more generally, backgrounds with a covariantly constant null Killing vector
and  flat or `conformal' transverse part, see,
e.g. \SANTAa\SANTAb\garriga\SANTAc\tsecov\Duval\ber)
and the backgrounds obtained from  gauged WZW models
(see e.g. \BCR\WIT\BSthree\HOR\BShet\ginspqu\kounn\napwi\GER\KTtwo).
Only the  latter class of solutions  has a direct CFT interpretation
(in terms of coset $G/H$ models \bha\GKO).
It was recently  understood  \NW\ that the two classes  actually  intersect.
 As a result,   some representatives of the first  class  can be  given a CFT
interpretation by identifying them
with members  of the second class  related to   non-semi-simple  groups
\NW\KK\etc\ORS\edc\sfedual.\foot{
The basic  solution of \NW\ is the WZW action for the non-semi-simple group
$E^c_2$ -- a central extension of the Eucledian group in two dimensions
\saletan\jao. Further studies of this model can be found in \KK\etc\KTone.
Larger classes of WZW models based on non-semi-simple groups were
constructed  in \ORS\ using a contraction method and  directly in \edc.
The explicit form of the action of these models can be found in \sfedual\
(see also \edc).
A class of coset models obtained by a particular gauging
of the general model of \ORS\ and equivalent to non-abelian duals
\nadual\ of WZW actions was  found in \sfedual.}

A subclass of  backgrounds with  a covariantly constant null Killing vector
and flat transverse part  (plane wave backgrounds) can be  described by the
following $\s$-model action ($i,j=1,2,\dots ,N$)
\eqn\one{\eqalign{S(u,v,x) =  & {1 \over \pi \a' } \ints  \ [  2\del u \bd v \
+ \
 ({ g}_{ij } + b_{ij})(u) \ \del x^i\bd x^j\ ] \cr  &-   {1 \over 8 \pi  }
\ints   \sqrt \g R^{(2)} \P(u) \ , \cr } }
where the functions $g_{ij}, b_{ij}, \P$ are subject to the equation \KTone
\eqn\plol{  g^{ij}  {\ddot g}_{ij} -\ha  g^{ij} g^{mn}\dot g_{im}\dot g_{jn}
 + \ha g^{ij} g^{mn}{\dot b}_{im}{\dot b}_{jn} + 2 \ddot \Phi =0\ , }
which expresses the condition of conformal invariance of \one\ to all orders
in $\a'$.\foot{The exactness  of this condition is easy to understand by noting
that higher-order corrections should depend on $\a'$ while  $\a'$ in \one\ can
be given an arbitrary value by rescaling   $v$ and $x^i$, i.e. by a coordinate
transformation. That implies, in particular, that the central charge of the
model should be equal to its  free-theory value $c= D=N+2$. }

Since one is ultimately interested in {\it four} dimensional solutions
it is important to  find  which of the  exact $D=N+2=4$ backgrounds
\one,\plol\
(in addition to the  already known $D=4$ model of \NW)  admit  direct  CFT
interpretation in terms of  $G/H$ cosets.\foot{One can prove either by a direct
computation \sfetse\ or by
using a theorem in \figsonia\ that  no WZW model based on a non-simple
group  \NW\noumo,
other than the one based on $E^c_2$ \NW, or its analytic continuations,
can be constructed in four
dimensions.} This is one of the  aims of the present paper.

The idea we shall use is the following (see also \etc).
We shall start with  the known $D=4$ solutions  corresponding to gauged
WZW models and having the structure\foot{We are assuming that there are  two
commuting isometries (shifts of $y^i$).
A number of such models are known (e.g.  cosets of products  of $SL(2, \IR)$
and $SU(2)$, see e.g. \HOR\kounn\napwi\GER ) and we will mention them in the
appropriate
parts of the paper.}
\eqn\act{\eqalign{S(t,r,y) =&{1\ov \pi \a'} \ints
[ - \del t \bd t +  \del r \bd r  +  \bl(G_{ij}+B_{ij}\br)(t,r) \del y^i\bd
y^j]\cr
&- {1\ov 8\pi} \ints \sqrt{\g} R^{(2)} \Phi'(t,r) , \cr} }
and identify the models in the class \one\ which are related to \act\
by performing a singular coordinate transformation and a rescaling of $\a'$
\eqn\red{ r  = \e\ v + u\ ,\qq  t  =u\ ,\qq
y^i= \sqrt \e \ x^i \ , \qq \a'\to \e\ \a'\ , \qq \e \ra 0 \  , }
\eqn\redd{
 (G_{ij}+B_{ij}\br)(u,u)= (g_{ij}+b_{ij}) (u) \ , \ \  \Phi'(u,u)= \P (u) \ .
}
 Since the $u,v$-term in the resulting action \one\ is boost-invariant, we can
also introduce a free parameter  by  rescaling  $u$  in $g_{ij}, b_{ij}, \P$.
We can also make a translation of $u$ by a constant
in one of two arguments of $G,B,\Phi'$.
As discussed in Appendix A  similar   limiting  procedure can be  applied to  a
more general class of $\s$-models (an example of such  model will be considered
in Section 5).

If the background fields in \act\ satisfy the string equations (which are
covariant relations)
the background fields in \one\ obtained by the  (singular) coordinate
transformation
\red,\redd\  are  guaranteed to satisfy the same equations, i.e. the conformal
invariance condition \plol.
Note that to get the exact expressions for $g,b,\P$ it is enough to start with
just  the leading-order (one-loop) expressions for $G,B,\P'$: $\a'$-dependent
terms in $G,B,\P'$ present in the `conformal' or `CFT' scheme (the one in which
the tachyon equation has  the standard Klein-Gordon form with no
$\a'$-corrections, see e.g. \DVV\BSexa\SFTS)  in any case would
drop out in the limit $\e\ra 0$ in \red.
Let us mention  also that the $O(2,2)$ duality rotations of the models \act\
(in the directions of the isometries $y^i$)
directly correspond to the duality rotations of the models \one\ \KTone.

Having identified  a model \one\ which originates from a model \act\ we can now
reconstruct a (non-semi-simple) coset  CFT   which is behind \one\
by taking the corresponding limit in  the (semi-simple) coset CFT corresponding
to
\act.\foot{The limit in \red\  with $r,t,y$ identified with coset coordinates
implies the corresponding limit in the coset current algebra construction
(i.e. in the OPE's or in the Lie algebra relations) and vice versa.
For the WZW models of \ORS\ and the coset models of \sfedual\ such a relation
has been found in  \sfedual.}

The examples of $D=4$ plane-wave backgrounds which admit an explicit  coset CFT
interpretation will be given in Sections 2 and 3.  All of them can be
constructed either directly  (by various gaugings based on the  $E^c_2$ WZW
model of \NW)
or by using the above limiting prescription.
For simplicity, in some cases we shall start  directly
 from an non-semi-simple coset CFT and only mention the original model which
gives it in the singular  limit, whereas
in other cases  we  shall  obtain the result via the limiting method and only
mention the underlying CFT.

In Section 4  we shall study  a number of   $D\leq 6 $ plane wave solutions in
bosonic string theory which have  explicit  coset  CFT counterparts.  Since
`extra' dimensions  will be  compact and  isometric
 the corresponding backgrounds  can be  given a  Kaluza-Klein  interpretation
as $D\leq 4$  `plane wave  with
gauge field'  solutions.
Some of these   bosonic backgrounds
    can  be  also identified with    $D\leq 4$ plane wave  solutions in  the
heterotic string  theory.  We shall  discuss   a number of examples
of such heterotic string solutions  which are closely related  to  $D=4$
black hole solutions
(for a previous discussion of plane wave solutions in the heterotic string
theory see  \SANTAa\ber).
 In many cases  the  prescription  \red\ can be applied
to obtain a plane wave solution from  a `cosmological' solution which
is related  by analytic continuation to   the   part of a black hole solution
inside  the horizon.
As an example,  we  shall  also  consider   a
  $D=4$  bosonic plane wave  related to  a bosonic  $D=4$  `charged black hole'
  background (with non-trivial axionic and electric charges)
based on the  coset
$[SU(2)\otimes SL(2,\IR) \otimes U(1)]/[U(1)\otimes U(1)]\  $ \ger.

In Section 5  we shall present another $D=4$
 solution which  also admits an exact CFT
description.
 It belongs to the
general class of backgrounds  with a covariantly constant null Killing vector
but  is not a plane wave.
It can be obtained by a non-abelian duality transformation  \nadual\ (see also
\GRV\grnonab\AABL\sfedual\AAL)  applied to the
$SU(2)$ WZW model and a singular limit, or directly
by the non-abelian duality transformation of the $E^c_2$  WZW model of \NW.

Some concluding remarks will be made in Section 6.
 A generalisation of the limiting procedure \red,\redd\
leading to backgrounds with  a covariantly constant null Killing vector will
be given in Appendix A, where,  as  an example,
  we  present  a five dimensional plane wave
solution related to the Schwarzchild black hole.
In Appendix B we shall discuss a CFT  description
of a  background which is dual to the flat $D=2$ space and which represents
the `transverse' part of the solution of Section 5.
In Appendix C we shall   show that the  charged black string model  \HOHO\
can be interpreted as  a $D=3$ bosonic model corresponding to the  $D=2$
charged black hole solution  \napp\ of the heterotic theory. This result will
be used in Section 4.


\newsec{Simplest  examples: $D=3$ gauged WZW and  $D=4$ WZW models}
\subsec{$D=3$ coset models}

To illustrate the above limiting prescription \red\   let us first consider the
 $D=3$ backgrounds (which form an obvious  subset of $D=4$ models with one
`spectator' dimension).
 Let us start with the  flat space model \act\ in cylindrical coordinates
\eqn\actt{S(t,r,y) ={1\ov \pi \a'} \ints\
[\   - \del t \bd t +  \del r \bd r  +  r^2 \del y\bd y\ ]
- {1\ov 8\pi} \ints\ \sqrt{\g} R^{(2)} \Phi_0 \ . }
The singular boost \red\ gives the following flat background (called `null
orbifold' in \HORS) \eqn\onee{S(u,v,x) ={1\ov \pi \a'} \ints\
[\   2 \del u \bd v  +  u^2 \del x\bd x\ ]
- {1\ov 8\pi} \ints\ \sqrt{\g} R^{(2)} \Phi_0 \ . }
If we start with the leading-order dual to \actt
\eqn\acttt{S(t,r,y) ={1\ov \pi \a'} \ints\
[\   - \del t \bd t +  \del r \bd r  +  r^{-2} \del y\bd y\ ]
- {1\ov 8\pi} \ints\ \sqrt{\g} R^{(2)} (\Phi_0 +2 \ln r) \  }
we get  another (now curved)  background in the class \one\ which is the exact
dual image \KTone\ of \onee
 \eqn\oneee{S(u,v,x) ={1\ov \pi \a'} \ints\
[\   2 \del u \bd v  +  u^{-2} \del x\bd x\ ]
- {1\ov 8\pi} \ints\ \sqrt{\g} R^{(2)} (\Phi_0 + 2 \ln u) \ . }
These models  can be considered as limiting cases of the
`neutral black string'   one.
If we start with the $SL(2,\IR)/U(1)\otimes \IR$ model in a particular
coordinate region
\eqn\blh{\eqalign{S(t,r,y) =&{1\ov \pi \a'} \ints
[ - \del t \bd t +  \del r \bd r  +  a^{-2} \tanh^2 ar\ \del y\bd y ] \cr &
- {1\ov 8\pi} \ints \sqrt{\g} R^{(2)} (\Phi_0  + 2\ln \cosh ar) \ \cr } }
which in the  formal limit $a\ra 0$  reduces to \actt\ (note that the zero
central charge condition restricts  the value of $a$: $\a'a^2 = (26-D)/6$)
we find
\eqn\blhh{\eqalign{S(u,v,x) =&{1\ov \pi \a'} \ints\
[\ 2\del u \bd v  +  a^{-2} \tanh^2 a u \ \del x\bd x\ ]\cr &
- {1\ov 8\pi} \ints\ \sqrt{\g} R^{(2)} (\Phi_0  + 2\ln \cosh au) \   . \cr } }
Here $a$ is a true free parameter (it can be absorbed into a rescaling of $u,v$
and $x$). The $a\ra 0$ limit of this model is the `null orbifold' \onee.

Starting with the $SL(2,\IR)/U(1)\otimes \IR$ model in the dual to \blh\ region
gives the model dual to \blhh\ with $a\inv\tanh \ra a\coth,  \cosh \ra \sinh$
(i.e. with $au \ra au + i\pi/2$)
which includes \oneee\ as
the $a\ra 0$ limit. If instead of $SL(2,\IR)$ cosets we start  with $SU(2)$
ones
we find \blhh\ with  hyperbolic functions replaced by trigonometric ones, i.e.
with $a\ra ia$.
The model \blhh\ was obtained in \KK\
by direct gauging  of the four-dimensional $E^c_2$  WZW model of \NW.

We conclude that the  $D=3$ plane wave type models
which have coset CFT interpretation are represented by
\eqn\bhh{S(u,v,x) ={1\ov \pi \a'} \ints\
[\ 2\del u \bd v  +  g(u) \del x\bd x\ ]
- {1\ov 8\pi} \ints\ \sqrt{\g} R^{(2)} [\Phi_0  + 2 \ln f(u) ] \  , }
where  the functions $g(u)$ and $f(u)$ can take the following  pairs of values
 \eqn\hhh{\eqalign{&g(u) = 1\ , \ \ u^2\  ,  \ \  \tanh^2 u \ , \ \  \tan^2 u \
, \ \
 u^{-2}\ , \ \ \coth^2 u\ , \ \ \cot^2 u \ , \cr
 &f(u) = 1\ , \ \ 1\ , \ \  \cosh u\ , \ \ \cos u \ , \ \
u\ , \ \ \sinh u \ , \ \ \sin u \ . \cr } }
The first two cases here correspond to flat spaces.
The free parameter $a$  can be re-introduced by shifting $u\ra au$ in \hhh.
Applying the limiting procedure similar to \red\ directly to $SL(2,\IR)$ WZW
model
does not give a new $D=3$ model: one finds a  background with a degenerate
metric (i.e. effectively a two-dimensional  space). This is consistent with the
absence of  non-semi-simple  $D=3$ Lie algebras with a  non-degenerate
invariant bilinear form \sfetse\figsonia.

\subsec{$D=4$ WZW models}
Let us now turn to $D=4$ models. It was shown in \etc\ that the first
non-trivial example \NW\ can be obtained
by taking the singular limit \red\ of the WZW action for
$SU(2)_k \otimes \IR$.
Indeed,  if we parametrize the $SU(2)$ group element as
\eqn\pasu{g= e^{i {\s_1\ov 2} \th_L}\ e^{i {\s_3\ov 2} \phi}\
e^{i {\s_1\ov 2} \th_R}  }
and the translational factor $\IR$ in terms of the time-like
coordinate $t$, then the WZW action is given by  (we omit the constant dilaton
term)
\eqn\acts{S={k\ov 4\pi} \int d^2 z\ \bl( -\del t \bd t + \del \phi \bd \phi +
\del \th_L \bd \th_L
+ \del \th_R \bd \th_R  + 2 \cos \phi\ \del\th_L \bd \th_R
\br) \ .}
This action has the same form as \act\ with $r=\phi , \  y^i=(\th_L,\th_R), \
\a'= 4/k$.  After making the redefinition  \red,\redd\ and taking  the limit
$\e\to 0$ the action \acts\ becomes identical to the
action for the $E^c_2$ WZW model \NW \ ($\a'=2$)
\eqn\acd{I_0 (g) ={1\ov 2\pi} \int d^2 z\
[\  2 \del v\bd u + \del x_1 \bd x_1
+\del x_2 \bd x_2 + 2 \cos u\ \del x_1 \bd x_2\ ] \ ,}
which is a particular member of the class \one.

Similar model (related by analytic continuation $v\ra -iv, u\ra iu$)
is obtained by starting with the WZW action for $SL(2,\IR)_k\otimes U(1)$
where the translational factor now is space-like (coordinate $r$) and the role
of time is played   by the non-compact coordinate ($i\phi$) in the
analog of \pasu\ for $SL(2,\IR)$. We  finish with  \acd\
with $\cos u$ replaced by $\cosh u$, and the corresponding symmetry group is
$P^c_2$ (i.e. the extension of the Poincar\'e group in two dimensions).
Now, however, there is an extra time-like direction in the $(x_1,x_2)$ plane
so that this model is not interesting by itself (but it is useful
as a starting point for constructing cosets, see subsec. 3.1).
Let us also note that \acd\ can be obtained by starting with the
$SL(2,\IR)_{-k}\otimes U(1)$ action in a particular patch (i.e.  with  \acts\
with
$\th_L\to i\th_L$, $\th_R\to i\th_R$ and $k\to -k$).

Other $D=4$ models of the type \one\  can be  found  by  applying
abelian $O(2,2)$ duality rotations  to the   $SU(2)_k \otimes \IR$
(or  $SL(2,\IR)_{-k}\otimes U(1)$)  WZW model  and taking the singular limit
\red.
The resulting models will be duality rotations of \acd.
Since the abelian duality transformations  of a  WZW model for a group $G$
can be represented  \dualmargi\KIRd\GiKi\  as  $(G\otimes H)/H$ gauged WZW
models
we are guaranteed to have a CFT description for the  resulting  backgrounds.
Other non-trivial examples  \KTone\ are
 obtained by starting from the $O(2,2)$ duals of $SU(2)$ or $SL(2,\IR)$ WZW
models equivalent to  $[SU(2)\otimes U(1)]/U(1)$
or $[SL(2,\IR)\otimes \IR]/\IR$ gauged WZW models (the latter is the charged
black string model of \HOHO).  We shall discuss the  CFT interpretation of the
corresponding plane-wave solutions explicitly  in the next section.

We can also start with a product of two  $D=2$ cosets
$G/H \otimes G'/H'$ (where $G,G'= SL(2,\IR) $ or $ SU(2)$ and $H,H'= \IR$ or
$U(1)$)
with signature $(-+++)$, i.e. with the model of the type  \act.
More general models correspond to its    $O(2,2)$ dual versions  (see, e.g.,
\HOR\GER\napwi\givpassq\gpr), i.e. the  `twisted'
gauged WZW models for $(G\otimes G')/H\otimes H'$.
The resulting models \one\ will also be discussed in Section 3.

Finally, there is a possibility to first
 apply the  non-abelian duality  transformation \nadual\grnonab\ to $SU(2)$ or
$SL(2,\IR)$ WZW models and
then  consider the limit analogous to \red\ (see  Appendix A). The resulting
model
will have a covariantly constant null Killing vector but will not belong to
 the class \one\  and will be discussed in detail in Section 5.


\newsec{$D=4$ plane wave solutions
corresponding to   gauged WZW models}

In this section we shall give explicit examples of four dimensional
plane-wave type
solutions with a particular emphasis on their description in terms of
 CFT's.
All of them  will correspond to gauged WZW models based on non-semi-simple
groups.

\subsec{$D=4$ plane-wave solution related to `black string'  background }

Let start with the current algebra for the direct product theory
$SU(2)_k \otimes U(1)\otimes U'(1)$.
We will perform a contraction \ORS\ with respect to the generator
$J_3\in su(2)$ and $J_0'\in u'(1)$ and gauge the abelian symmetry generated by
the linear combination $J_1\in su(2)$ and $J_0\in u(1)$.
The result of the contraction is
the direct product of the the current algebra for $E^c_2$ \NW\ and a
$U(1)$ factor
\eqn\opec{\eqalign{&J P_i \sim {i \e_{ij} P_j \ov z-w}\ ,
\qq P_i P_j\sim {i\e_{ij} F \ov z-w} + {\d_{ij}\ov (z-w)^2} \qq (i=1,2) \cr
& J F\sim {1\ov (z-w)^2}\ ,\qq P_0 P_0 \sim {1\ov (z-w)^2}\ .\cr}}
Let us consider now the coset theory $[E^c_2 \otimes U(1)]/U(1)$  with the
stress tensor  given by
\eqn\stres{
T=\  \ha :(P_i^2 +2 J F - F^2):\  +\ \ha {:P_0^2:} \
- \  {:(P_1 + q_0 P_0)^2:\ov 2(1+q_0^2) }\ ,}
where the constant $q_0$ parametrizes the embedding of $H=U(1)$ into
$G=E^c_2 \otimes U(1)$.

In order to find the $\s$-model corresponding to the above coset CFT we
parametrize the group element $g\in G$ as
\eqn\parae{g=\diag(g_1,g_2)\ ,
\quad g_1=e^{i x_1P_1} e^{ i u J} e^{i x_2 P_1} e^{i v F}\in E^c_2\ ,\quad
g_2=e^{i \phi P_0}\in U(1)\ .}
The action of  the $[E^c_2 \otimes U(1)]/U(1)$ gauged WZW model  reads
\eqn\act{\eqalign{S = & I_0(g_1)
+{1\ov \pi} \ints\ [\ \ha \del\phi \bd\phi
+i A\ ( \sqrt{q_0}\ \bd \phi + \bd x_1 + \cos u\ \bd x_2)\cr
&- i \A\ ( \sqrt{q_0}\ \del \phi+ \del x_2 + \cos u\ \del x_1)
+ A\A\ (1+ \cos u + 2 q_0)\ ]\ ,\cr } }
where the action $I_0(g_1)$ for the $E^c_2$ WZW model is given in \acd.
The gauge transformations that leave \act\ invariant are
\eqn\gaue{ \d x_1=\d x_2=\n\ ,\quad \d u=\d v=0\ ,
\quad \d \phi=2 \sqrt{q_0}\ \n
\ ,\quad \d A=i \del \n\ , \quad \d \A= - i \bd \n\ .}
After fixing the gauge $\phi=0$, integrating over the gauge fields and
changing the variables
\eqn\chvar{x_1\ra \ha( { x_1 \over \sqrt{ q_0}} + {x_2\over \sqrt{1+q_0}} \ )\
,\qq
x_2\ra \ha ( { x_1\over \sqrt{ q_0}} - {x_2\over \sqrt{1+q_0}}\ )\ }
we obtain the  $\s$-model action \one\ with  the background fields given by (we
rescale $u\ra 2 u, \ v\ra v/2 $)
\eqn\exo{\eqalign{&ds^2=2 dvdu + {\cos^2 u\ov q_0+\cos^2 u}\ dx_1^2 +
{\sin^2 u\ov q_0 + \cos^2 u}\ dx_2^2  \cr
&b_{12} = {\sqrt{q_0(q_0+1)}\ov q_0+\cos^2 u}\cr
&\Phi= \ln (q_0+\cos^2 u) + \const  \ .\cr}}
For $q_0=0$ the above solution reduces to the direct product of the coset model
$E^c_2/U(1)$ of \KK\  and a free boson, or, equivalently,  to a particular case
of \bhh,\hhh\ with an extra free dimension.
The  background   \exo\  with $q_0\not=0$ can be  found  from the
latter one by  a duality transformation.

The solution  \exo\ can be  also obtained by applying the
limiting procedure of Section 1 to the direct product theory of
the `Eucledian' black string model \HOHO\ and a translation,
i.e. $[SU(2)\otimes U(1)]/U(1) \otimes \IR $.

The analytic continuation $u\to i u$, $v\to -i v$, $x_2\to i x_2$ and
$q_0\to -q_0$ is  changes $E^c_2$ into $ P^c_2$
and the compact $U(1)$ factors into the
 non-compact $SO(1,1)$ ones. The trigonometric functions are then replaced
by their hyperbolic counterparts  and the background
becomes asymptotically  (for large $u$)  Minkowski
with zero antisymmetric tensor and linear dilaton ($\P$ depends only on $u$ and
thus does not
contribute to the central charge).

A number of related solutions can be found by taking various  limits of the
parameters in \exo\ and making analytic continuations.
For example,  the  special limit of \exo\ : $q_0 =-1- q^2\e,\   u\to \sqrt{\e}\
u\ ,
 \ x_1\ra i\sqrt{\e}\ x_1, \ x_2\ra i x_2$ gives the following
interesting solution
\eqn\ext{\eqalign{&ds^2=  2dvdu+ {1\ov u^2+q^2}\
dx_1^2 + {u^2\ov u^2+q^2}\ dx_2^2
\cr
&b_{12}= { q \ov u^2+q^2}\cr
&\Phi=\ln (u^2 + q^2) + \const \ .\cr}}
It can be obtained by gauging other combinations of
currents in the direct product $E^c_2\otimes U(1)$, or by applying a
limiting procedure similar to \red\ to the case of the
direct product  of $E^c_2/U(1)$ model (see eq.(2.8) of \etc),
and a translation $\IR$, or by applying a duality (combined with a coordinate)
transformation to Minkowski space \etc.
%
The direct product models  based on  \bhh,\hhh\  with $g(u)=u^2$ or $u^{-2}$
are special limits of this solution.

\subsec{$D=4$ plane-wave solutions related to  twisted products of two $D=2$
 cosets  }
Let us start with the current algebra for the direct product
$SU(2)_k\otimes SL(2,\IR)_{-k'}$.
The OPE's for the currents $\{I_3,I_i\}\in su(2)_k$ $\ (i=1,2)$ are
\eqn\ssul{I_i I_j \sim {i\e_{ij} I_3\ov z-w} + {k \d_{ij}\ov 2(z-w)^2 }\ ,\qq
I_3 I_i \sim {i \e_{ij}I_j\ov z-w}\ ,\qq I_3 I_3 \sim {k\ov 2( z-w)^2 }\ ,}
and for $\{J_3,J_i\}\in sl(2,\IR)_{-k'}$ $\ (i=1,2) $ are
\eqn\ssu{J_i J_j \sim {i\e_{ij} J_3\ov z-w} + {k'\d_{ij}\ov 2(z-w)^2 }\ ,\qq
J_3 J_i \sim {- i \e_{ij}J_j\ov z-w}\ ,\qq J_3 J_3 \sim {-k'\ov 2(z-w)^2}\ .}
The stress energy tensor of the direct product theory and the corersponding
central charge are
\eqn\stre{T={:I_i^2 + I_3^2:\ov k+2} + {:J_i^2-J_3^2:\ov k'-2}\ ,
\qq c={3k\ov k+2}+{3k'\ov k'-2}\ .}
We shall do a contraction with
respect to the generators $I_3\in su(2)$ and $J'_3\in sl(2)$. We define
a new basis $\{I_3,I_i,J_3,J_i\}\to \{J,F,P^{\a}_i\}$, $(i,\a =1,2)$ as
\eqn\pp{\eqalign{
&J=I_3+ a J_3\ ,\quad F=\e(I_3- a J_3)\ ,\quad P_i^1=\sqrt{2\e}\ I_i\ ,
\quad P_i^2 = \sqrt{2\e}\ a \ J_i \cr
& k={1\ov \e}\ ,\quad k'={1\ov  a^2 \e}\ , \quad a^2 = {k\ov k'} \ , \cr }}
where $a$ is a free parameter.
Then in the singular limit the OPE's of the resulting six dimensional current
algebra (which will be denoted by $su(2)_2^c$) are
\eqn\opect{J P^{\a}_i \sim {i \e_{ij} \l_{\a} P^{\a}_j \ov z-w}\ ,
\ \   P^{\a}_i P^{\b}_j\sim {i\d_{\a\b}\e_{ij} \l_{\a} F \ov z-w}
+ {\d_{\a\b}\d_{ij}\ov (z-w)^2}\ ,\ \   J F\sim {1\ov (z-w)^2}\ , }
where $\l_{\a}=(1,a)$. The corresponding stress tensor and central charge are
\eqn\strec{T=\ha :\bl[P^{\a}_i P^{\a}_i +2 J F -(1+a^2) F^2 \br]:\ ,\qq c=6\ .}
Let us note that, our contraction,
though similar in spirit,   differs from that of \ORS. In particular,  notice
the
presence of the `contraction' parameter $a$.\foot
{In the case of $a=1$ (`diagonal' contraction) the algebra \opect\ can  be also
obtained by contracting the $so(4)$ algebra as in \ORS\
with respect to an $so(2)$ subalgebra and neglecting one free field
(see \sfedual, App.C).
In the notation of \sfedual\ $su(2)^c_2\simeq so^*(4)^c_{so(2)}$.}
To find the corresponding  $\s$-model we parametrize the group element as
\eqn\parr{g=e^{i P_1^{\a} x_1^{\a}} e^{i u J} e^{i x_2^{\a} P_1^{\a}}
e^{i v F}\ .}
Then the WZW action reads\foot{
Throughout this paper we  shall use the notation $u'= a\ u+ d$, where $a$ and
$d$ are constant free parameters.
This action  is  conformal for arbitrary values  of $a$ and $d$,
being a solution of \plol.
 While the origin of $a$ is clear from the current algebra \pp\  this is not so
for the parameter $d$.  In fact, when
$d\neq 2\pi n$, $n\in Z$ the corresponding CFT is not the current algebra
\opect\ for $SU^c_2$.  To understand the origin of the free parameter
$d$  one should start with
 the eight dimensional WZW action for the direct product
$E^c_2\otimes E^c_2$ and gauge the subgroup generated by $F_1+F_2$.
Because this subgroup is nilpotent there will be a constraint condition to be
solved (besides the usual gauge fixing).
Its most general solution will contain a parameter  related to $d$.}
\eqn\accd{I_0(g)={1\ov 2\pi} \int d^2 z\
[\  2 \del v\bd u + \del x_1^{\a} \bd x_1^{\a}
+\del x_2^{\a} \bd x_2^{\a} + 2 \cos u\ \del x_1^1 \bd x_2^1
+ 2\cos u'\ \del x_1^2 \bd x_2^2\ ]\ .}
To obtain a four-dimensional model we should gauge a two-dimensional subgroup.
We choose to gauge a symmetry that acts in a left-right asymmetric fashion
\BSthree\ on the group element, i.e.
$\d g= i\n_{\a}( t_l^{\a} g - g t^{\a}_r)$, $\n_{\a}=(\n_1,\n_2)$, where
\eqn\symet{\eqalign{&t^1_l= P^1_1 + q P^2_1\ ,\qq t^2_l= - q P^1_1 +P^2_1 \ ,
\cr
& t^1_r = P^1_1 - q P^2_1 \ ,\qq t^2_r = q P^1_1 +P^2_1\ ,\cr }}
with $q$ being a `twisting' parameter.
Then the gauged WZW action becomes (we suppress the indices $(\a,\b)$)
\eqn\acctt{\eqalign{ S=& I_0(g) + {1\ov \pi} \ints
[ iA \bl(m_1(q) \bd x_1 +m_2(q) \bd x_2\br)\cr
&- i\A \bl(m_2(-q) \del x_1 +  m_1(-q) \del x_2 \br)  + A M \A ]\cr }}
\eqn\acct{\eqalign{& m_1(q)= \pmatrix{1& q \cr -q & 1}\ ,
\qq m_2(q)=\pmatrix{\cos u & q\cos u'\cr -q \cos u & \cos u'} \cr
& M=\pmatrix{ (\cos u-1) -q^2 (\cos u' +1) & q (\cos u+\cos u')\cr
- q (\cos u+\cos u')  & (\cos u'-1)  - q^2 (\cos u+1) }\ .\cr }}
The gauge transformations that leave \acct\ invariant are
$$\d x_1 = -m_1(-q) \n \ ,\quad \d x_2 = m_1(q) \n\ ,
\quad \d u=\d v =0\ ,
\quad  \d A= -i \del \n\ ,\quad \d \A= - i \bd \n \  . $$
Fixing the gauge as $x_2^{\a}=0$ and integrating out the gauge fields gives
(we change  the coordinates $x_1^{\a}\to  m_1(q) x_1^{\a}$, rename
$x^{\a}_1\to x_{\a}$ and rescale $u\to 2u$, $v\to v/2$)
\eqn\exf{\eqalign{
&ds^2= 2dvdu\  + \ {  \D}^{-1} \ (\cos^2 u \sin^2 u'\ dx_1^2 +
\cos^2 u' \sin^2 u \ dx_2^2 )\cr
&b_{12}= {q}\ {  \D}^{-1} \ \cos^2 u \cos^2 u'\ ,
\qq \Phi = \ln \D  + \const \cr
& \D \equiv    \sin^2 u \sin^2 u'  +   q^2 \cos^2 u \cos^2 u'   \ , \ \
u'\equiv au + d  \ .   \cr } }
Notice that for zero twisting $q=0$ \exf\ corresponds to a plane wave which
can be obtained from
to the direct product $SU(2)/U(1)\otimes SL(2,\IR)/\IR$  \  \kounn\  by
performing  the singular coordinate transformation \red,\redd.
The  solution  \exf\ can be obtained  from  that  direct product  model
by  performing a one-parameter $O(2,2)$ duality transformation,
or by  `boosting' according to  \red\
the backgrounds of \HOR\napwi\givpassq\GER. Note that the solution \exo\ can be
obtained from \exf\ after we rescale
$x_i\to \sqrt{2(q^2-1)}\ x_i$, set $d=\pi/4$, $q_0=1/(q^2-1)$
 and take the limit $a\to 0$.

The background \exf\ and its various analytic continuations and limits
can be represented also in the following general form
\eqn\duali{\eqalign{&ds^2= 2dudv +  {g_1 (u') \ov g_1 (u')g_2 (u)  + q^2 }\
dx_1^2 +
 {g_2 (u)\ov g_1 (u)g_2(u')  + q^2} \ dx_2^2  \cr
&b_{12 }= {q \ov  g_1 (u') g_2 (u) + q^2}   \cr
&\Phi= \ln \bl(f_1^2 (u')f_2^2 (u) [ g_1(u') g_2(u) + q^2]\br) + \const \ , \cr
}}
where the functions $g_i, f_i$ can take any pairs of values in \hhh\ (\duali\
is related by \red\ to
a similar background in  \givpassq).
In particular,  the solution \exo\ is a special case of \duali\ with
$g_1=1, \ g_2=\tan^2 u$.
Also,  \ext\ is a  special   case of \duali\ with $g_1=1, \ g_2 =u^2$,
i.e.  the  background \ext\ is  actually  dual to the flat space.



\newsec{Bosonic  and   heterotic  plane wave solutions
with  abelian gauge fields}

\lref \barss {I. Bars, ``Curved spacetime geometry for strings and affine
non-compact algebras", USC-93/HEP-B3, hep-th/9309042.}

 In this  section  we  shall consider a number of   $D\leq 6 $ plane wave
solutions in bosonic string theory which have   explicit  coset  CFT
counterparts.  If `extra' dimensions are  compact and represent isometric
directions
 the corresponding backgrounds  can be  given a    Kaluza-Klein  interpretation
 as $D\leq 4$  plane waves with extra
gauge fields.
Some of these  solutions    can  be  also identified with    $D\leq 4$ plane
wave  solutions in  the  heterotic string  theory.
Since  the special  heterotic  plane wave  backgrounds  we shall find below
will
be related to the exact bosonic  plane wave solutions, they will  represent the
exact  (all order in $\a'$) solutions of the heterotic string theory.
In addition to the obvious $(1,1)$ supersymmetric versions of the solutions of
the previous sections we shall  find  non-trivial `asymmetric' solutions with
$(1,0)$
world sheet supersymmetry.
%

\subsec{General remarks}
As in Sections 2,3  one can construct  bosonic plane wave solutions
by starting with the known $G/H$ gauged WZW models  and applying
the limiting procedure  \red,\redd.
If one is interested in the  dimensions   of the resulting  \sm configuration
space being $D=5,6$ (with one time-like direction)  a few simplest
choices  for $G/H$ are again  products of cosets
of $\sll,\  \su $ and  $\uo$'s and  various `twisted' gaugings  (duality
rotations), e.g.,  for
$D=5:\ \   \sll/\uo \otimes \su \  , \ \  \bl[\sll\otimes \su \otimes
U(1)\br]/[U(1)\otimes U(1)]\ ,  ... \ ,  \ \ $ and for
$\ D=6:\ \  \sll\otimes \su \ , \ \ \bl[\sll\otimes \su \otimes U(1)\br]/U(1)\
, \ \
 \bl[\sll\otimes \su \otimes U(1)\otimes \uo \br]/[U(1)\otimes U(1)] \ ,  ...
\ . $
The  conformal  \sms  obtained from such  gauged WZW models can be represented
as
\eqn\het{\eqalign{S=&{1\ov \pi \a'} \ints
[  \bl(G_{\m\n}+B_{\m\n}\br)(x) \del x^\m\bd x^\n + \cA_{n\m} (x) \del x^\m \bd
Z^n        + \bA_{n\m }(x) \bd x^\m \del Z^n\cr
&+  G_{mn} (x) \del Z^m \bd Z^n \ ]
- {1\ov 8\pi} \ints \sqrt{\g} R^{(2)} \Phi(x) \ ,\cr }  }
 i.e. in the form of  an   action of a
bosonic string  propagating in  $D=4+ r$  space-time $(x^\m, Z^m)$
($\m,\n=0,1,2,3$,  $m,n=1, ...,r$) with $G_{\m n}=\ha ( \cA_{n\m} +\bA_{n\m} )$
       and $B_{\m n }=  \ha ( \cA_{n\m} - \bA_{n\m})$.
Since the background fields do not depend on the  coordinates $Z^n$  one  may
 interpret  such  a  model
 in a Kaluza-Klein  manner as a $D=4$  background with  non-trivial  gauge
(and scalar) fields.  One can represent \het\
in the form  which  is invariant under the
the space-time gauge transformations of $\cA^{n(+)}_{\m}, \ \cA^{(-)}_{n\m}$
(see below)  if one also   shifts  $Z^n$ and  makes the  `anomalous'
transformations of  $B_{\m\n}$
\eqn\hett{\eqalign{S=&{1\ov \pi \a'} \ints
[  \bl(G^{(4)}_{\m\n}+B_{\m\n}\br)(x) \del x^\m\bd x^\n + Z^n {\cal F}_{n\m\n}
(x) \del x^\m\bd x^\n  \cr
&+ G_{nm}(x) (\del Z^n  +   \cA^{n(+)}_{\m} (x) \del x^\m)(\bd Z^m +
  \cA^{m(+)}_{\m} (x) \bd x^\m) ]\cr
&- {1\ov 8\pi} \ints \sqrt{\g} R^{(2)} \Phi(x) ,\cr }  }
where
 $$\cA^{(\pm)}_{n\m} \equiv   \ha ( \cA_{n\m} \pm \bA_{n\m}) \ , \ \ \
\cA^{m(+)}_{\m} \equiv  G^{mn}\cA^{(+)}_{n\m} \ , \ \ \      {\cal
F}_{n\m\n}\equiv \del_\m \cA^{(-)}_{n\n}
 - \del_\n \cA^{(-)}_{n\m} $$
 and
\eqn\mett{G^{(4)}_{\m\n}= G_{\m\n}  -  G^{mn} \cA^{(+)}_{m\m}\cA^{(+)}_{n\n} }
is the  gauge-invariant  $D=4$ `Kaluza-Klein' metric.

One can also  try  to interpret   \het\  (e.g.,  with $\bA_\m=0$) as  a
bosonic part of the heterotic string action  coupled   to $\cA_{n\m}$  with
$Z^n$   being  now    chiral scalars representing the `right' internal  sector
of the heterotic string (see, e.g., \GR\gpr). In this case, however,
it is not a priori clear why the full action (with
the  additional  `left' fermionic terms  implied by (1,0) supersymmetry) should
also be conformally invariant, i.e.  should
represent a solution of the  heterotic string theory.

In general, there are several ways to construct heterotic string solutions
related to gauged WZW models. One can start with a bosonic coset, consider it
formally as a solution of the  superstring theory (conformal (1,1)
supersymmetric  model)
and embed this superstring solution into the (1,1) supersymmetric subset of
solutions of the heterotic string theory by adding  the gauge field background
equal to the (generalised) Lorentz connection (see \GRT\ and refs. there).  In
this way one  cancels the $2d$ gauge anomaly  which otherwise makes direct
(1,0) supersymmetric version of $G/H$  gauged WZW model inconsistent \GRT\ (for
a discussion of a potential unitarity problem and some earlier refs. see also
\barss).
In  particular,  to  construct the heterotic string analog of the  $\sll/\uo$
`black hole'  solution  \WIT\ of the  bosonic string theory
 one  needs   to add  a  $U(1)$  gauge field background \GRT\
\eqn\hebh{ ds^2= dr^2 + a^{-2} \tanh^2 ar \ dy^2 \ , \ \  \Phi= \Phi_0 + \ln
\cosh^2 ar
\ , \ \  \  \cA_y=- {1 \over \cosh^2 ar } \ . }
This  solution  can be related  to
the  `charged $D=2$ black hole' (or `neutral  black string') solution of \ISH\
(the bosonised form of the heterotic string action can be identified with the
model \het\ where   $D=3, \ r=1, \ G_{mn} =1 $ and  $ \bA_\m = - \cA_\m$ so
that the Kaluza-Klein redefinition of the metric \mett\ is trivial, cf. (C.1)
with
$q=0$).

For example,  the bosonic plane wave solutions
of Sects. 2,3  are obviously also the solutions
of the superstring theory and as such can be embedded into the set  of
solutions
of the heterotic string theory by supplementing them  with  gauge field
backgrounds equal to the values of the  Lorentz connection.

Another possibility is to consider directly the (1,0) supersymmetric  extension
of the  group $G$ WZW theory   or, more generally, of the chiral gauged  $G/H$
WZW theory  \KSTh\  (with the dimension of the configuration space
being still equal to dim $G$ and thus the $2d$ anomaly being
harmless).\foot{The $G/H$ chiral gauged WZW model with an abelian subgroup $H$
is equivalent
to a particular $(G\otimes H)/H$ gauged WZW model, or to a particular duality
rotation of $G$ WZW model \KSTh. Analogous relation is probably true  also in
the case of a non-abelian $H$. }
One can  also  construct `heterotic cosets' by starting with the (1,0)
supersymmetric version of  anomalously gauged  (1,0) supersymmetric  WZW model
and adding    internal `right' fermions in order to cancel the
 total $2d$ gauge anomaly \john.
Once bosonised, the total heterotic string action  should be equivalent to a
particular bosonic anomaly-free gauged WZW model \GPS\john.\foot{The bosonic
version of the action of
`$S^2$ plus monopole' solution of \GPS\
can  thus  be  essentially identified \GPS\john\ with the action of $SU(2)$ WZW
model (the original suggestion to reinterpret
a group  $G$ WZW action as a sigma model on $G/H$ with an $H$- gauge field
background was made in \absa, see also \khou).
For other examples see  \john\nels. In particular,
the  bosonic action of the $[SL(2, \IR)\otimes SU(2)]/[U(1)\otimes U(1)]$
heterotic  model of \john\ can be also  considered
 as the action of  a
$\bl[SL(2, \IR)\otimes SU(2)\otimes U(1)\otimes U(1)\br]/[U(1)\otimes U(1)]$
bosonic gauged WZW model.}

Thus in order to interpret a particular $bosonic$ $G/H$  coset
as  representing   a $heterotic$ string solution one should be able to
identify
the  corresponding sigma model \het\ with the full  bosonised  action
of the heterotic string in  a  given  background \GPS\john\ (for some related
earlier refs. see \duff).
Namely, \hett\ should be equivalent  to a bosonised version of   a  special
case of the
heterotic string  action \hulw\
\eqn\hettt{\eqalign{S=&{1\ov \pi \a'} \ints
\bl[  \bl(G^{(4)}_{\m\n}+B_{\m\n}\br)(x) \del x^\m\bd x^\n  +
i \psi_a [\delta^a_b \bd  + \omega^a_{b \m} (x) \bd x^\m] \psi^b \cr
& +i \Psi_I [\delta^I_J \del  + \cA^I_{J \m} (x) \del x^\m] \Psi^J
 +  {\cal F}^I_{J ab }(x) \Psi^J\Psi_I \psi^a\psi^b \ \br] \ .
\cr  } }
The $D=4$ heterotic string background is then  given  by $G^{(4)}_{\m\n}, \
B_{\m\n}, \  \cA_{n\m}$ and $\P$. Note that the scalar background $G_{mn}(x)$
(which is present in the corresponding bosonic Kaluza-Klein solution)  is
absent  in the heterotic action; it appears only after \hettt\ is bosonised and
put in the form of \het\ (see also \john).


\subsec{Heterotic  string plane wave solutions }

Let us start with  $D=3$ examples, first reviewing   how
one can represent  the $SU(2)$ WZW action  in the form \het, i.e.
as a  \sm on $S^2$ with   a $U(1)$ monopole coupling (see \absa\GPS\john).
Using the parametrisation \pasu\ one finds (cf. \het,\mett)
\eqn\acth{\eqalign{S&={k\ov 4\pi} \int d^2 z\ \bl(  \del \phi \bd \phi + \del
\th_L \bd \th_L
+ \del \th_R \bd \th_R  + 2 \cos \phi\ \del\th_L \bd \th_R
\br) \cr
&={k\ov \pi} \int d^2 z\ \bl(  \del \phi' \bd \phi' +
  \sin^2 \phi' \del \vp \bd \vp
+ \del Z' \bd Z'   - 2 \sin^2 \phi' \  \del\vp \bd Z'
\br) \cr
&={k\ov 4\pi} \int d^2 z\ \bl[  \del \phi \bd \phi +
  (\sin^2 \phi  +  \fourth \cA_\vp \cA_\vp )  \del \vp \bd \vp
+ \del Z \bd Z    + \cA_\vp \del\vp \bd Z
\br]  }}
where
$ \  \phi=2\phi' \ , \ \ \vp= \th_L \ , \ \  Z=2Z'= \th_L + \th_R \  $ and
\eqn\monop{ ds^2 = d\phi^2 + \ \sin^ 2 \phi \ dx^2 \ , \ \ \ \Phi=\Phi_0 \ , \
\ \   \cA_x = 2(\cos{\phi }-1) \ . }

We are now  able to interpret the  $E^c_2$ WZW model \acd\ of \NW\
as a  $D=3$  heterotic string plane wave solution with a vector field.\foot{
Since the internal coordinate should be  periodic  one  should also factorize
over a discrete subgroup, as in \GPS. The same remark holds for all
models in this section.}
In fact,   adding an extra time direction $t$ as in \acts\  and combining it
with $\phi$  according to  \red\
we get  another equivalent representation  for \acd\ (cf. \het,\mett)
\eqn\acthh{ S= {k\ov 4\pi} \int d^2 z\ \bl[ 2 \del v \bd u  +
  (\sin^2 u  +  \fourth \cA_x \cA_x )  \del x \bd x
+ \del Y \bd Y   + \cA_x (u)  \del x \bd Y
\br]  \ , }
i.e. the following $D=3$ background
\eqn\threeh{ ds^2 = 2dvdu\  + \ \sin^ 2 u \ dx^2 \ , \ \ \ \Phi=\Phi_0 \ , \ \
\   \cA_x = 2(\cos{u }-1) \ . }
Analytic continuation gives  related solution with $\sin \to -\sinh, \ \cos\to
\cosh$ (which  is found  directly if one replaces the monopole  model \acth\ by
the $D=2$
anti de Sitter one  obtained from $\sll$ WZW  theory  \lost).
We also get a  $D=3$ heterotic plane wave solution
by `boosting' (cf. \blh,\blhh)  the  product of the  $D=2$ black hole solution
\hebh\ with  a time line.

Another  $D=3$ plane wave solution is obtained by starting with   the  $D=2$
`charged black hole'
heterotic string solution  (the solution \hebh\ is related to a special case of
this background) \napp\
\eqn\yos{ ds^2 =f^{-1} dr^2  -f dy^2  \ , \ \  \ \  f= 1- 2m{\rm e}^{-Qr} + p^2
{\rm e}^{-2Qr} \ , } $$    \  \  \Phi=\Phi_0 +  Q r \ , \ \  \ \  \cA_y = - \
\sqrt 2 p \  {\rm e}^{-Qr}
 \  .   $$
It can be  given  a     world sheet  interpretation  in terms of the $D=3$
bosonic conformal \sm  \het\  by starting with the action of the  $D=3$
charged black string  ($[\sll\otimes \uo]/\uo$ coset) model \HOHO\ or by
constructing
$\sll$ `heterotic' cosets \john\ (related  background  with a non-constant
scalar field
can be interpreted  a  $D=2$ bosonic Kaluza-Klein background \lost).
Considering the region between the horizons and singularity
one can  represent  \yos\ as a cosmological solution \tset \
\eqn\yyo{ ds^2 = -dt^2 + \   g(t) \  dx^2 \  ,\ \ \ \
g(t) =  {   {\rm cosh}^2 bt\  {\rm sinh}^2 bt \ov ( {\rm cosh}^2 bt + \g  \
{\rm sinh}^2 bt)^2} \ ,  }
$$ \Phi (t)  = \Phi'_0 + {\ln }( {\rm cosh}^2  bt
+ \g \ {\rm sinh}^2 bt) \  ,    \ \ {\cal A}_{x} (t)
 = - { \sqrt {-2 \g}  \ov  (\g + 1)   \ ({\rm cosh}^2  bt
+ \g \ {\rm sinh}^2 bt ) } \ ,   $$
where $x=i(1+ \g) y ,  \ Q=-2ib, \ p^2/m^2  = - 4  \g /( 1- \g )^2.
$
Taking special limits and making analytic continuations  one can relate \yyo\
to the two other $D=2$ heterotic solutions  \hebh\ and \monop\ discussed above.
The background \hebh\ is reproduced  when  $ p= 0, \ Q= \sqrt{2/ \g}, \ b=ia $
and $\g\to 0$ (comparing \yyo\ with \hebh\ one needs to take into account a
rescaling of the gauge potential due to a  rescaling of coordinates).
Another limit $\g\to -1$  leads to the solution
related to \monop\ by  $t\to i \phi$ (assuming one absorbs an infinite constant
in the
gauge potential  into a gauge transformation and rescales $k$ in \acth).
This relation is explained in Appendix C where we show that the
$D=3$ model  representing the bosonised form of   the heterotic solution \yyo\
is equivalent to the $D=3$ charged black string  \HOHO\ or $[\sll\otimes
\uo]/\uo$ gauged WZW model.

Adding a spatial
direction  to \yyo\ and applying  \red\ we finish with the $D=3$ heterotic
plane wave
\eqn\yyoy{ ds^2 = 2dvdu  +   g(u) \  dx^2 \  , \ \  \Phi =\Phi (u)\ , \ \ {\cal
A}_{x} = {\cal A}_{x} (u) \ ,  }
which includes  \threeh\
 as a particular case.
Given the relation  of Appendix C between \yyo\ and the charged black string
background
 one can also relate the corresponding plane wave solutions \yyoy\
and \exo\ in a similar way  (namely,  \exo\ is
 the $D=4$ bosonic background  which is the bosonized version of the $D=3$
heterotic solution \yyoy).

We can now   construct $D=4$ heterotic plane waves by
starting with the  direct products of $D=2$ heterotic solutions
\hebh,\monop,\yyo.
Since,   as was noted above, \yyo\  is the most general solution, we get
the most general `direct product' $D=4$ model by tensoring  the two models
\yyo.
  In view of the result of the Appendix C the  corresponding  $D=6$ bosonic
model is
equivalent (up to analytic continuations)
to the product of   $[\sll\otimes \uo]/\uo$   and $[SU(2)\otimes \uo]/\uo$
models with
 different values of  `mixing' parameters $\g_1, \ \g_2$ and   levels $k',\  k$
(or parameters  $b',\ b$).
The  plane wave  solution obtained  after applying  \red\ to  such product
 will be (cf. \duali)
\eqn\yyy{ ds^2 = 2dvdu   + g_1(u') \  dx^2_1  +   g_2(u) \  dx^2_2 \  , \ \
\Phi = \Phi_1 (u') +\Phi_2 (u)\ , } $$ {\cal A}_{1} = {\cal A}_{x} (u') \ ,  \
\  \ \ {\cal A}_{1} = {\cal A}_{x} (u) \ ,  \ \  \ u'=au + d\ ,\  \ a^2= k/k' \
,  $$
where the functions $g, \Phi, \cA_x$
 are the same as in \yyo\ with  $bt\to u$, $\g\to \g_i$.

Let us  consider explicitly some particular cases of \yyy.
 For example,  a  `throat' limit of the  $D=4$ black hole solution of \GM\
is described by the product of the $D=2$ black hole \hebh\ and
the monopole theory \acth\  \GPS\  (note that the gauge field background in
\hebh\
can be ignored  in the leading order approximation since it can be considered
as being higher order in $\a'$  but is essential in order to have an exact
heterotic string solution \GRT).
The resulting background  (equivalent to  the  $D=6$ bosonic plane wave
solution obtained from the coset   $[SL(2, \IR)_{-k'}\otimes  U(1) ]/
U(1)\otimes  SU(2)_{k}$, i.e. from the direct product of the model of \ISH\ and
 $\su$ WZW model)
can be  represented as
a $D=4$  heterotic  plane   wave  (cf.  \exf,\duali,\yyy )
\eqn\hetpl{\eqalign{&ds^2 = 2dvdu\  + \ \tanh^2 u'\ dx^2_1 \  + \  \sin^ 2 u \
dx^2_2 \ , \ \ \ \Phi=\Phi_0  + \ln \cosh^2 u' \cr
 &\cA_1=- {1 \over \cosh^2 u' } \ , \ \  \   \cA_2 = 2(\cos{u }-1) \ . } }
The $D=6$   bosonic plane wave  (with the action $S'$
 is equivalent to \accd) which
 one obtains  from the $\sll_{- k'}\otimes \su_{k}  $ WZW
model\eqn\slllu{\eqalign{S=&{k'\ov 4\pi} \int d^2 z\ \bl[ - \del \phi \bd \phi
+ \del \th_L \bd \th_L
+ \del \th_R \bd \th_R  + 2 \cos \phi\ \del\th_L \bd \th_R
\br) \cr
& + a^2 (  \del \phi' \bd \phi' + \del \th'_L \bd \th'_L
+ \del \th'_R \bd \th'_R  + 2 \cos \phi'\ \del\th'_L \bd \th'_R) \br]\ , \cr }}
\eqn\sllluu{\eqalign{S'=  &{1\ov \pi\a'} \int d^2 z\ \bl[ 2\del v \bd u + \del
x_1 \bd x_1
+ \del x_2\bd x_2  +
\del x_3 \bd x_3
+ \del x_4\bd x_4  \cr
&+ 2 \cos  u' \  \del x_1  \bd x_2  +  2 \cos  u\ \del x_3  \bd x_4 ]
 \ ,  }}
corresponds to the following $D=4$ heterotic string solution
\eqn\bert{\eqalign{&ds^2 = 2dvdu\  + \ \sin^2 u'\ dx^2_1 \  + \  \sin^ 2 u \
dx^2_2 \ , \ \ \ \ \ \ \Phi=\Phi_0  \cr
 &\cA_1= 2(\cos{u' }-1)\ , \ \  \   \cA_2 = 2(\cos{u }-1) \ .
 } }
This background  can be  also related (by analytic continuation and boost) to
the Bertotti-Robinson solution of \lost, i.e. to the direct product of the
`anti de Sitter' and `monopole' $D=2$ solutions.

Another  special  case of \yyy\ is obtained by starting with the direct product
of \yos\ and \acth\ (which is related to the bosonic solution    describing  a
`throat' limit of the electrically charged  $D=4$ black hole of \lost).
The resulting background is an  obvious  superposition of \threeh\ and \yyoy.

 It is also possible to  go beyond the direct product models \yyy.
One can, in fact,  construct various  conformal  $D=6$ bosonic cosets by
applying
the   $O(4,4)$ duality rotations to  \slllu,\sllluu\ (cf. \duali).
It is not clear a priori   which  of them can be re-interpreted as $D=4$ {\it
heterotic}
string solutions (duality may not `commute' with bosonisation/fermionisation).
An  example  of a non-direct-product model where this is possible was given in
\john.
It is straightforward to write down a plane wave  background which is  related
by \red\ to the $D=4$  `dyonic black hole' heterotic string  solution of \john.
Equivalent $D=6$ bosonic plane wave  is obtained from a particular
 $\bl[SL(2, \IR)\otimes SU(2)\otimes U(1)\otimes U(1)\br]/[U(1)\otimes U(1)]$
bosonic gauged WZW model.

\subsec{ A $D=4$ bosonic plane wave background with   $U(1)$ gauge field}
As an example of a  less trivial bosonic
 model which is not just a direct product of lower dimensional ones
let us consider  the plane wave analog of the  bosonic
`charged black hole' solution of \ger\
 based on  the  bosonic $\bl[SL(2,\IR)_{-k}\otimes SU(2)_{k'} \otimes
U(1)\br]/[U(1)\otimes U(1)] $  gauged WZW model.
The corresponding \sm is given by \het\  with  $D=5$ (i.e. $r=1$), $\ G_{mn} =1
, \  \cA_\m \not=0, \  $ and $
\bA_\m=0 $.  It  can  be interpreted in a  Kaluza-Klein  manner  as a $D=4$
bosonic background with an extra gauge field. Note that   the Kaluza-Klein
redefinition of the metric \mett\   here is  non-trivial (this redefinition was
ignored in \ger\  so the  analysis of  properties of the resulting $D=4$
background is to be reconsidered).
In contrast to  the bosonic backgrounds of the previous subsection this
background does not  at the same time represent  a $D=4$ solution of the
heterotic string theory.

The non-vanishing  $D=4$ components of the  (unredefined)  metric $G_{\m\n}$,
antisymmetric tensor, dilaton  and   vector potential  in \het\ are\foot{We
make  a particular choice of possible free parameters with  $Q$ being a real
number (related to the electric and axionic charges \GERexa). We correct two
erroneous coefficients (in front of  $B_{t\phi}$ and  $A_\phi$) in the final
expressions of \ger.}
\eqn\blg{\eqalign{ds^2&=-{\sinh^2\rho\ (1+Q^2\sin^2\theta)\over \cosh^2
\rho+Q^2\sin^2\theta}\ dt^2+
a^2{d\rho^2 } +d\theta^2 + {\cosh^2 \rho\  \sin^2\theta\over
\cosh^2\rho+Q^2\sin^2\theta}\
d\phi^2\cr
&=-{(r-1)(1+Q^2\sin^2\theta)\over r+Q^2\sin^2\theta}\ dt^2+
{a^2dr^2\over 4 r(r-1)} + d\theta^2 + {r\sin^2\theta\over r+Q^2\sin^2\theta}\
d\phi^2  \cr
&B_{t\phi}=Q\ {(r-1)\sin^2\theta\over r+Q^2\sin^2\theta} \ , \ \  \
\Phi=\ln(r+Q^2\sin^2\theta) + \const \cr
&\cA_t= 2Q\sqrt{Q^2+1}\
{(r-1)\sin^2\theta\over r+Q^2\sin^2\theta} \ , \ \ \ \
\cA_{\phi}=-2\sqrt{Q^2+1}\
{r\sin^2\theta\over r+Q^2\sin^2\theta} \  ,  }}
where $a^2= k/k'$ and   $r\equiv \cosh^2 \rho >1$.
The proper  gauge-invariant $D=4$  metric  $G^{(4)}_{\m\n}$   is given by
\mett.
The $Q=0$ limit of the model
is equivalent to the direct product of the  $D=2$ black hole
and  `$S^2$ plus monopole'   model \acth\
and thus is the same as  the   limiting case  of the $D=4$  black solution
of \GM. The absence of an  extra gauge field component (cf. \hebh)
implies that \blg\  is $not$  an exact  solution of the heterotic string
theory.

 If we formally set (cf. \red, \redd)\foot{As in the case of all other
black-hole type  backgrounds  to obtain a plane wave
 we  actually consider  the region below the horizon, i.e.  we first make the
analytic continuation $\rho \ra i \rho$.  In this case  \blg\ takes the form of
a cosmological metric:
 $t$  becomes space-like and  $\rho$ plays the role of a time coordinate.}
\eqn\chvo{\rho= iu\ ,\ \  \th=\e a\inv  v + au +  d \ ,\ \ t= - \sqrt{\e}\ x_1\
,
\ \ \phi=\sqrt{\e}\ x_2\ , \  \  Z\to \sqrt{\e}\ Z\ ,\ \  \a'\to \e\ \a'\ ,}
and take the limit $\e\to 0$  we obtain  the following plane wave  background
with $Q$, $a$ and $d$ as free parameters ($u'\equiv  au + d$)
\eqn\solu{\eqalign{
&ds^2= 2dvdu + {\sin^2 u\  (1+ Q^2 \sin^2 u') \ov \cos^2 u + Q^2 \sin^2u' }\
dx_1^2
+ {\cos^2 u \  \sin^2 u'\ov \cos^2 u + Q^2 \sin^2 u'}\ dx_2^2  \cr
&b_{12}=  Q\ {\sin^2 u \sin^2 u' \ov \cos^2 u + Q^2\sin^2 u' }\ , \ \ \ \Phi=
\ln(\cos^2 u + Q^2 \sin^2 u')  + \const \cr
&\cA_1= 2Q\sqrt{Q^2+1} \
{\sin^2 u \sin^2 u'\ov \cos^2 u + Q^2 \sin^2 u'} \ , \ \  \ \cA_2
=-2\sqrt{Q^2+1}\ { \cos^2 u \sin^2 u'\ov \cos^2 u + Q^2 \sin^2 u'} \ . }}
Similar backgrounds are obtained by analytic continuations in $Q, a, d$ and the
coordinates.
 It is straightforward  to write down  explicitly the  resulting
world-sheet action \het\  as a sigma model \one\  corresponding to  a $D=5$
plane-wave solution and to check that the  all-order conformal invariance
condition
\plol\ is satisfied.
The CFT  description of  this  solution  is  based on the coset
$\bl[SU(2)^c_2 \otimes U(1)\br]/[U(1)\otimes U(1)]$
which has $c=5$, cf. Sect.3.2.

Computing the  $D=4$ metric $G^{(4)}_{\m\n}$  \mett\
 of this bosonic `plane wave with a gauge field'  background we get
$$ (ds^2)^{(4)}= 2dvdu  +     {\sin^2 u [\cos^2 u ( 1 + Q^4 \sin^4 u')
 +  Q^2\sin^2 u'( 1 + \cos^2 u -  \sin^2 u \  \sin^2 u') ] \ov (\cos^2 u + Q^2
\sin^2u')^2 }\ dx_1^2$$ $$ +  {\cos^2 u  \sin^2 u'( \cos^2 u \cos^2 u'  +
Q^2\sin^2 u \sin^2 u') \ov \ (\cos^2 u + Q^2 \sin^2 u')^2}\ dx_2^2 $$ $$ +
2Q(Q^2 + 1)  {\sin^2 u \cos^2 u  \sin^4 u' \ov  (\cos^2 u + Q^2 \sin^2 u')^2}\
dx_1 dx_2  \ .  $$


\newsec{Solution with  a covariantly constant null Killing vector: non-abelian
dual of $E^c_2$ WZW model}

Let us consider the non-abelian duality transformation on the WZW
action for $E^c_2$ \acd\ with respect to the group  $E^c_2$  itself.
We start with the action \nadual
\eqn\sdual{\eqalign{
S_{\rm dual}= &I_0(g) + {1\ov \pi} \ints\ \Tr\bl[ A \bd g g\inv - \A g\inv \del
g +
A g \A g\inv - A\A \cr & -2 i \l (\del \A - \bd A -[A,\A] ) \br]\ ,}}
where the group element $g\in E^c_2$ is parametrized as in \parae.
In the $E^c_2$- Lie algebra basis $\{P_i,J,F\}$
the Lagrange multipliers and
the gauge fields are   $\l=\{\l_i, \l_3,\l_4\}, \ $ $A=\{ A_i, A_3, A_4\}$.
{}From the infinitesimal gauge transformations
$\d g= [g,i \n]$, $ \d \l=[\l,i \n]$, where $\n=\{\n_i,\n_3,\n_4\}$
one finds
\eqn\infnt{\eqalign{&\d x_1 + \d x_2= \tan \u\ [ 2 \n_2 +  (x_2-x_1)\ \n_3]\ ,
\ \  \d x_1 - \d x_2 = -2 \n_1 +  (x_1+ x_2)\ \cot \u \ \n_3   \cr
&\d v= -(x_1+x_2)\ \n_2  + x_1 (x_1 + x_2\ \cos u)\ \n_3\ ,\qq      \d u=0 \cr
& \d \l_i=\e_{ij} (\l_3 \n_j - \n_3 \l_j)\ ,\qq \d \l_3=0\ ,
\qq \d \l_4= -\e_{ij} \l_i \n_j\ . } }
 The corresponding transformations for the gauge fields  are
\eqn\infgf{\d A_i=-i \del \n_i + \e_{ij}(A_3 \n_j -\n_3 A_j) \ ,\qq
\d A_3=-i\del \n_3\ ,\qq \d A_4= -i \del \n_4 - \e_{ij} A_i \n_j\ ,}
and similarly for $\A$.
A convenient gauge choice is $x_1=x_2=\l_2=0$.\foot{
This choice introduces
a non-trivial Faddeev-Popov factor $\sim \l_1 \tan \u$ in the
path integral measure. It can be verified using the appropriate
expressions below that when this factor is combined with the measure of
the original WZW model (in our case this is the Haar measure  $\sim \sin u$)
it gives $e^{\Phi} \sqrt{-G}$, as expected \KIR\BSthree.}
In this gauge the action \sdual\ takes the following explicit form (we shift
$\l_4 \to \l_4 + v$)
\eqn\sdgf{\eqalign{&S={1\ov \pi}\ints\ [\ \del v\bd u - i A_1 \bd \l_1
- iA_3 \bd \l_4 + i A_4(\bd u - \bd \l_3)
+i\A_1 \del \l_1 + i \A_3 \del \l_4   \cr
& -i \A_4 (\del u- \del \l_3)
+  \bl( -2 \sin^2 \u\ \d_{ij} + (\sin u-\l_3)\ \e_{ij} \br) A_i\A_j
-\l_1 (A_2 \A_3 - A_3 \A_2 ) \ ]\ .\cr } }
After integrating over the gauge fields\foot{Note that the fields $A_4$, $\A_4$
appear in \sdgf\ only linearly, imposing the $\delta$-function condition $\l_3
= u + d $, where $d=\const$. This should have been expected since
we could have gauge fixed  only three parameters even though the gauge group
$E^c_2$
is a four dimensional one (notice that $\l_3$ is inert
under the gauge transformation \infnt).  } we obtain a $\s$-model  action
with  the following  couplings (we introduce new coordinates $x_1,x_2$ with
$x_1=\l_1$ and $x_2= -\l_4$)
\eqn\exff{\eqalign{& ds^2= 2dvdu + {1\ov x_1^2 \sin^2 {u\ov 2}}\
[\ 4 \sin^4 {u\ov 2}\ dx_2^2 + \bl(x_1 dx_1 +(\sin u -u-d )\ dx_2\br)^2 \ ]
\cr
&\Phi=\ln(x_1^2 \sin^2 {u\ov 2}) + \const \ , \ \ \ \ B_{\m\n}=0 \ . \cr}}
This   solution, though not a plane wave,   belongs to  a general
class of  backgrounds   with a covariantly constant null Killing vector
\Brinkman\tsecov. While the abelian duality transformations
do not lead us out of the  class \one\ of plane wave backgrounds \KTone\
this is no longer true for the non-abelian duality.
The solution \exff\  can be also obtained by applying the limiting procedure of
Appendix A to the
direct product of the time line and  the background of the
 non-abelian dual to the $SU(2)$ WZW action
(see eqs. (6.10),(6.11) of \grnonab).

By construction, \exff\  must  satisfy the one-loop  conformal invariance
 condition
but is not an exact solution  to all orders in the $\a'$-
expansion in the  standard `conformal' scheme.
The condition of conformal invariance of a background
 with a covariantly constant null Killing vector and $v$-independent dilaton
 is that the  `transverse' parts of its fields  (obtained by setting
$u=\const$)
should also  represent  a conformal theory, i.e. should satisfy the conformal
invariance  equations to all orders \tsecov.
In the present case  \exff\  ($a,b=\const$)
\eqn\tran{
ds_{\perp}^2 = {1\ov x_1^2} [4 a^2 dx_2^2 +{1\ov a^2}(x_1dx_1 + b \ dx_2)^2] \
, \ \ \  \Phi_{\perp}=\ln x_1^2 + \const \ . }
 By a shift $x_2\to c_1 \ x_2 + c_2 \ x_1^2$ and  rescaling of  $x_1$
 \tran\ can be put into the form
\eqn\traa{ds_{\perp}^2= dx_1^2 + {1\ov x_1^2}\ dx_2^2 \ , \ \ \ \
\Phi_{\perp}=\ln x_1^2  + \const \ ,  }
which is one-loop conformally invariant,  being dual to the flat $D=2$
Eucledian space.\foot{The general solution of the one-loop conformal invariance
equations  in
two dimensions in the conformal gauge $ds^2=\Om^{-1} dzd\z$, $\Phi=\ln \Om$
has $\Om=A z\z + B z + C \z + D$. The equation for the central charge
gives $c=2 + \d c$, with
$\d c\sim A$. If  $A\neq 0$ then  shifting the
coordinates $z$, $\z$ we can set the constants $B$ and $C$ equal to zero.
This solution
is described by the $SU(2)/U(1)$ CFT. If we demand that $\d c=0$ then
by a change of coordinates we obtain \traa.
The CFT description of \traa\ is given by the coset $E^c_2/[\IR \otimes U(1)]$
 \etc.
The all-order solution that corresponds to a correlated
limit taken in the exact background for the coset $SU(2)/U(1)$ \DVV\ can  be
found in \etc\ (see also Appendix B).} However, it receives $\a'$-corrections
in the standard `conformal' scheme (in which there are $\a'$-corrections the
$SL(2, \IR)/U(1)$ Euclidean  $D=2$ black hole background. As for the $D=2$
black hole
(and for the $D=3$ black string and, most probably, for all gauged WZW models)
\tpl\SFTS\ there exists a special scheme
in which the leading-order solution \traa\ (and thus  also \exff)  is  actually
the exact solution to all orders in $\a'$.

The CFT description of \exff\ is based on  the coset $G^c_g/G$ with
$G=E^c_2$.\foot{More generally, according to \sfedual\ the  non-abelian duality
transformations of  the WZW action for a group $G\otimes U(1)^{dim(G/H)}$
are equivalent to gauged WZW models $G^c_h/H$ where by $G^c_h$ we denoted the
non-semi-simple group obtained by contracting the direct product group
$G\otimes H$ as in \ORS, and where we gauge the diagonal subgroup.}
The presence of the  $\a'$-corrections to the {\it non-abelian}  dual \exff\ in
the standard `conformal' scheme is  consistent  with
the
coset description \sfedual\   and is not in conflict with the result of \KTone\
(which  only states
that the {\t abelian}   $O(d,d)$ duality
transformations performed on exact plane wave
solutions \one,\plol\  give exact  solutions belonging to the same class \one).

\newsec{ Concluding remarks }
In this paper we have  presented a number of $D=4$ plane-wave-type
bosonic and heterotic string solutions  which admit explicit coset CFT
description, thus supplementing the  previously discussed  $D=4$  example \NW.
The knowledge of  underlying CFT is very important in order to
study the properties of these backgrounds (see \KK).
For example, the conformal  algebra fixes the form
of the equations for the string modes. We find,  for example,  that
 the tachyonic equation does not receive $\a'$ corrections in  the standard
`CFT' scheme.\foot{In general, the  tachyon equation on a plane-wave background
may contain
$\a'$ corrections since the argument that a rescaling of $\a'$ in \one\ is
equivalent to a rescaling of $x^i$ does not apply once an $x$-dependent tachyon
coupling $ T(x)$ is added to the action.}
For the plane wave backgrounds \one\ the Klein-Gordon-type  equation
\eqn\klgor{ - {1\ov e^{\Phi} \sqrt{-G}}\ \partial_{\m} \bigl(
e^{\Phi}\sqrt{-G}\ G^{\m\n}\
\partial_{\n} \bigr)\ \Psi(x)= E \Psi(x)\ ,}
can be solved explicitly  (see also \klim\garriga\jonu)
\eqn\klsol{
\Psi_{p,k_i}(u,v,x^i)=C \bl( e^{\Phi}\sqrt{g}\ \br)^{-1/2}\ e^{\ha i p v}\ e^{i
k_i x_i}\
e^{{i\ov p}[Eu - \int du g^{ij}(u) k_i k_j]}\ .}
Let us mention also that it is possible to  find explicitly
the bosonisation rules  and free field description for all of our
examples since they all are related in one way or another to cosets of
$SU(2)_k$.
The bosonisation rules for
$E^c_2$  (see \KK)  can be very easily derived from the known ones for
$SU(2)_k \otimes \IR$ (however,  a  relation of the corresponding
representation theories
is unclear).

\bs\bs

\centerline{ {\bf Note added } }

While  this paper  was in preparation
there appeared  an interesting  preprint  \antob\ where  some $D=4$ bosonic
plane-wave type backgrounds with a
covariantly constant null Killing vector  were  given  a CFT interpretation
in terms of the  coset  model $(E^c_2 \otimes E^c_2)/E^c_2$.  From the point of
view
of the present paper
such  solutions  should correspond to a
 limit (similar to  \red) of  the backgrounds  related to
the coset $[SU(2)\otimes SU(2)\otimes \IR^2 ]/[SU(2)\otimes \IR]$.
The bosonic solutions of our Section 3 are  different  from the solutions
of \antob. At the same time, our
 solution  \exff\ can be obtained from
eq.(5.10) of \antob\ by taking the limit, $\kappa=1+\e $, $\e \to 0$
and  rescaling  the   coordinates.
This could  be expected in view of  the  general result of \sfedual\  about  a
relation between non-abelian duals and singular limits of some special cosets
(see footnote 16).


\appendix A {Limiting procedure  for a  general class of $\s$-models}

The discussion of a  limiting procedure  \red\ in Section 1  can be repeated
for a more general class of $\s$-models with the aim to obtain  other
backgrounds with a covariantly constant null Killing vector.
Consider  the  $\s$-model    action
\eqn\apa{\eqalign{ S=&{1\ov \pi \a'} \ints\
[\ \bl(G_{ij}+B_{ij}\br)(\phi,y)\ \del y^i\bd y^j
+ \bl(G_{ab}+B_{ab}\br)(\phi,y)\ \del \phi^a \bd \phi^b
\cr
& +A_{ia}(\phi,y)\ \del y^i \bd \phi^a + B_{ai}(\phi,y)\ \del \phi^a \bd y^i\ ]
- {1\ov 8\pi} \ints\ \sqrt{\g} R^{(2)} \Phi(\phi,y)\ ,   \cr }}
where we split $D$ coordinates into two sets $\phi^a$ ($a=1,2$) and $y^i$
($i=1,...,N$) and make the coordinate transformation $\{\phi^a, y^i \}\to
\{u,v,x^i\}$
\eqn\appa{\phi^1 = \e\ v + u\ , \qq \phi^2=u\ ,\qq
y^i= f^i(\e;x)\ , \qq \a'\to \e\ \a'\ . }
The set of functions $\{f^i(\e;x)\}$ must be
chosen in such a way that the limit $\e\to 0$ is well defined and the new
action corresponds to a string background with a covariantly constant null
Killing vector
\eqn\apap{\eqalign{
S=&{1\ov \pi\a'}
\ints\ [ \ 2 \del v\bd u + (g_{ij}+b_{ij})(u,x)\ \del x^i\bd x^j +
a_i(u,x)\ \del x^i \bd u \cr &+ b_i(u,x)\ \del u \bd x^i \ ] \ - \ {1\ov 8\pi}
\ints\ \sqrt{\g} R^{(2)} \Phi(u,x)\ . }}
If the original action \apa\ is exactly conformal  to all orders in $\a'$ then
the resulting one \apap\ will be exactly conformal  as well. However, if the
original action represents  a leading order
 solution, then the fact that in the limit $\e\ra 0$ we have $\a'\ra 0$  does
not necessarily imply that
the resulting action will correspond to an   exact solution
to all orders in $\a'$
(note, in particular,  that  the inverse transformation to \appa\
is not defined in the limit $\e\to 0$).
Still, this is  the case when
the resulting background fields  depend
only on the light cone variable $u$ and not on $x^i$.
Since the central charge of the original model \apa\
is given by a power series
in $\a'$ the central charge of the new model \apap\
 will be $c=D=N+2$, i.e. will be equal to the number of degrees of freedom.
Such cases were considered in Sections 2,3,4. The background  from Section 5
which explicitly depends on $x^i$ is an example of a leading-order solution.

We can apply this procedure to other leading-order solutions with an idea to
get exact plane-wave  solutions or leading-order solutions with  covariantly
constant  null Killing  vector.\foot{A different singular coordinate
transformation
procedure for generating new string solutions  from known ones was discussed in
\KP.}
Consider, for example, the following class of metrics
\eqn\mee{ \a'  ds^2  = - f(r) dt^2  + h(r) dr^2 +  G_{ij} (r,t,y) dy^i dy^j \
}
and  change the coordinate $r\to z$ to put  \mee\ in the form
\eqn\meee{ \a'  ds^2  = F(z)(- dt^2  +  dz^2) +  G'_{ij} (z,t,y) dy^i dy^j \ .
}
If we now set $z= \e v + u \ ,\ \  t=u, $  rescale $\a'$ by $\e$,  assume that
there exists  an appropriate  redefinition of    $y^i$  (see  \appa)
and take the limit $\e\to 0$ we finish with ($\a'=1$)
\eqn\meet{  ds^2  = 2F(u)dudv  +  g_{ij} (u,x) dx^i dx^j \ ,  }
or a background with a null Killing vector
(further redefinition of $u$ puts it in the standard form).

Similar  procedure  can be applied to  various  black-hole type solutions
(cf. \yos, \yyo, \yyoy).  In the case of the $D=4 $ Schwarzchild black hole
it gives the rather trivial (flat)  background
$$   ds^2  = 2dudv  +  u^2 (dx_1^2 + dx_2^2)  $$
which is  an obvious solution of \plol\ and a special case of  \duali\ (with
$q=0$ and $g_i,f_i$
corresponding to  the second flat case in \hhh).
A Less trivial
 example is found if we start with   the five-dimensional solution that is the
direct product of the  $D=4$ Schwarzchild black hole with an additional
 space-like direction $y$
\eqn\sch{\a' ds^2 =  -(1-{m\ov r})\ dt^2 + (1-{m\ov r})^{-1}dr^2
+ r^2( d\th^2 + \sin^2\th \ d\phi^2) + dy^2 \ .}
 Replacing $r$ by a new coordinate $z$ defined by
(we consider the region inside the horizon)
$ z= - m(\psi + \ha \sin 2\psi)\ , \ r\equiv m \cos^2 \psi , $
we then set $$ \a'\to \e \a'\ ,\ \  \ z=- \e v + u \ ,\  \  \  y=u\ ,  $$ $$ \
 \ t=\sqrt{\e}\ x_1\ ,\ \  \
\th=\sqrt {\e}\ \r \ , \ \  \  x_2= m\rho \cos \phi\ , \  \ \  x_3= m\rho \sin
\phi\ , $$
take the limit $\e\ra 0$ (cf. \appa),  and define $\psi (u)$ as a solution of
$ u= - m(\psi + \ha \sin 2\psi)$. The result is
the following $D=5$ plane-wave background (which of course is a solution of
\plol)
\eqn\fdim{ ds^2 = 2dvdu + \tan^2\psi(u)\ dx_1^2 + \cos^4 \psi(u)\ (dx_2^2 +
dx_3^2)
\ .}


\appendix B {Coset CFT description of background  dual to  $D=2$ flat
space-time}

Let us consider the $SL(2,\IR)_{-k}/SO(1,1)$ current algebra theory and
perform the $\rm{In\ddot on \ddot u}$-Wigner-type contraction:  $J_{\pm}\to
P_{\pm}/\sqrt{\e}$, $J_0\to F/\e$ and $k\to k/\e$.
The resulting OPE's, stress tensor and central charge are
\eqn\tobe{\eqalign{&P_+ P_- \sim { F\ov z-w}+ {k\ov (z-w)^2}\ ,
\quad F P_{\pm}\sim 0\ ,\quad F F\sim 0 \cr
&T={1\ov 2k} :(P_+ P_- + P_- P_+ -{1\ov k} F^2):\ ,\qq c=2\ .\cr }}
The constant $k$  can be rescaled   by redefining
the current algebra generators but it is useful to keep it (note that we cannot
drop the $F^2$ term in the stress tensor since it is needed for the closure of
the Virasoro algebra).
 The only current that has a regular OPE with the stress tensor is
$F$ (in the original coset this  was the property of   $J_0$).
We can find the  $\s$-model corresponding to the stress tensor in \tobe\ using
the algebraic operator method \DVV\BSexa.
Representing  the zero modes of the currents as first order differential
operators
\eqn\zerom{ P_+ = -{1\ov a}(x_1 + x_2) {\del\over \del x_1} -{\del\over \del a}
\ ,\qq P_-= a {\del\over \del x_2}\ ,
\qq F= {\del\over \del x_1} - {\del\over \del x_2} \ ,}
and restricting \BSexa\ the $L_0$ (`Klein-Gordon') operator to states
satisfying ${\del\over \del a}\Psi=0$ we obtain the following $D=2$ metric and
dilaton \eqn\medi{\eqalign{&ds^2={1\ov (x_1+x_2)(x_1+x_2-2/k)}\bl[-{1\ov k}
(dx_1^2 + dx_2^2 ) +  2 (x_1+x_2-1/k)\ dx_1 dx_2 \br]  \cr
& \Phi=\ha \ln\bl[(x_1+x_2)(x_1+x_2-2/k)\br] + \const \ .\cr} }
One  can diagonalize the metric in the region $x_1+x_2 < 0$
(changing the coordinates
$x_1 = -{1\over 4} t^2 + \ha x\ , \ \  x_2=- {1\over 4}  t^2 - \ha x\ $)
\eqn\medii{ds^2 = -dt^2 + {1\ov t^2+ 4/k}\ dx^2\ , \ \ \ \ \  \Phi=\ha \ln
[t^2(t^2+4/k)\br] + \const \ .}
This metric has a cosmological interpretation of a Universe that starts from
a collapsed state at $t=-\infty$,  expands and reaches its maximum size at
$t=0$ and then recollapses at $t=\infty$ (see also \mull\cosm\TSEd).
As long as $k$ is finite,
there is no curvature singularity at $t=0$. This is reminiscent of the
analogous
situation for the $2D$ black hole \BSexa\PERRY. In the region where $x_1+x_2 >
0$
we analytically continue $t\to i\r$ and rename $x\to  t$. Then a curvature
singularity appears at $\r=\sqrt{4/k}$. This corresponds to a naked singularity
region of the $D=2$ black hole.
For  $k\to \infty$ the analytically continued version of \medii\ coincides
with \traa.
The background \medii\ is  thus dual to the $D=2$
flat space-time with a constant dilaton.

\appendix C {Relation between $D=3$ bosonic charged black string background
and
$D=2$ heterotic charged black hole  solution}
We would like to identify the $D=3$  \sm
corresponding to the  $[SL(2,\IR)\otimes \IR]/\IR$  gauged WZW model
(charged black string model \HOHO) with the bosonised  form  (see Section 4.1
and \GPS\john) of the heterotic string action in the charged black hole
background  \napp.
Namely, the aim will be to represent the former model in the `Kaluza-Klein'
form \het,\hett\ and to identify the  metric, gauge potential and dilaton of
\yyo\
with the corresponding couplings in \hett,\mett.

\def \t {\th}
\def \tt {\tilde \th}

The  action of the  $[SL(2,\IR)\otimes \IR]/\IR$  (axially)  gauged WZW model
 \HOHO\ can be represented in the form (see, e.g., \SFTS; we use the Euler
angle coordinates on $\sll$, $\t=\ha (\th_L-\th_R), \  \tt
= \ha (\th_L+\th_R) $ and fix the gauge where the extra $\IR$-field is zero)
$$ S(r, \t,\tt)= {k \over  \pi } \int d^2 z [   G_{\m\n }  \del x^\m \bd x^\n
+   B_{\t \tt}    ( \del \t \bd \tt  -  \del \tt \bd \t ) ] \ , \eq{C.1}
 $$
$$  ds^2 = G_{\m\n} dx^\m dx^\n=
\fourth dr^2 +  (1 + q)  {C-1 \ov C+ 1+2q } d\t^2
  -q {C+1 \ov C + 1+ 2q } d\tt^2 \  , \eq{C.2} $$
 $$   B_{\t \tt}   =  {q(C-1)\ov  C+1+2q}\ , \ \ \  \Phi=\Phi_0  +  \ln (C+1 +
2q)  \ ,   \ \ \ C\equiv \cosh r\ , \eq{C.3} $$
where $q$ is a free parameter. Changing the coordinates to $r, \t, Z$
$$  \t \to \t +  {q\ov q+1 } Z\  , \ \qq  \tt \to  Z \ ,  \eq{C.4} $$
we can re-write  the action (C.1)  as \het\ where $x^\m=(r, \t)$
and
$$  G_{\t\t}= (1 + q)  {C-1 \ov C+ 1+2q } \ ,
\ \  G_{ZZ}=  -  {q\ov q+1 }\ , \ \  \  $$ $$
\cA_{\t}=  {q\ov q+1 } G_{\t\t} + B_{\t \tt}  =  {2q(C-1)\ov  C+1+2q} \ , \
\ \  \ \bA_{\t}=  {q\ov q+1 } G_{\t\t} - B_{\t \tt} = 0 \ . \eq{C.5} $$
Shifting $ B_{\t \tt}$ by a constant leads to a constant shift of the gauge
potentials.  Computing the   gauge-invariant  $D=2$ `Kaluza-Klein' metric
\mett\ we find
$$  G^{(2)}_{\t\t} = G_{\t\t} - \fourth G_{ZZ}^{-1}\cA_{\t}^2 =
{(q+1)^2 (C^2-1)\ov  (C+1+2q)^2}  \  , \eq{C.6} $$
i.e. the following $D=2$ background
$$ ds^2 = \fourth dr^2  + {(q+1)^2 (\cosh^2 r -1)\ov  (\cosh r+1+2q)^2} d\t^2 \
, \ \ \ \ \Phi=\Phi_0  +  \ln (\cosh r+1 + 2q)  \ ,  \  $$ $$
\cA_{\t}=  {2q(C-1)\ov  C+1+2q} =  2q - {4q(q+1) \ov  C+1+2q}\ ,
\eq{C.7} $$
which becomes equivalent to  the analytic continuation of
\yos\ or \yyo\   with
$\g= -q/(q+1)$ after we   drop a constant in the gauge potential
and rescale the coordinates (a similar  to (C.7) form  of \yos\ appeared in
\john).  The background (C.7) was also
obtained in \john\ by a `heterotic' gauging of $\sll$.
The above discussion implies that the  model of \john\  when viewed as a
specific  conformal bosonic model is actually equivalent to the
$[SL(2,\IR)\otimes \IR]/\IR$  gauged WZW model.

The model (C.1)--(C.3) has two obvious limits: $q=0$ (neutral black string or
$SL(2,\IR)/\IR \otimes \IR$  gauged WZW model) and $q=\infty  $ ($\sll$ WZW
model).
This  explains  the observation in Sect.4.2
that  the   corresponding limits $\g=0$ and  $\g=-1$
of the  $D=2$ background  \yyo\
are  equivalent to the  exact $(1,1)$ supersymmetric
heterotic black hole solution \hebh\ (which, from the  bosonised
$D=3$ point of view is related to the model of \ISH\ or,  after change of
coordinates, to the neutral black string)
and the monopole model \monop\ (related to $SU(2)$ WZW theory).

\listrefs
\end